\documentclass[aps,prd,twocolumn,amsmath,amssymb,floatfix,nofootinbib,superscriptaddress]{revtex4-1}
\usepackage[utf8]{inputenc}
\usepackage{xcolor}
\usepackage{amsmath}
\usepackage{amssymb}
\usepackage{graphicx}
\usepackage{natbib}
\usepackage{hyperref}
\usepackage{aas_macros}
\usepackage[normalem]{ulem}

\newcommand{\utaumap}[1][]{\ensuremath{\tau^{#1 \mathrm{biased}}}}
\newcommand{\modmap}[1]{\ensuremath{#1^{\mathrm{mod}}}}
\newcommand{\nhat}[1][]{\ensuremath{\hat{\mathbf{n}}}}
\newcommand{\obsmap}[1]{\ensuremath{#1^{\mathrm{obs}}}}
\newcommand{\lenmap}[1]{\ensuremath{#1^{\mathrm{len}}}}
\newcommand{\unlmap}[1]{\ensuremath{#1^{\mathrm{prim}}}}
\newcommand{\biasmap}[1]{\ensuremath{#1^{\mathrm{biased}}}}
\newcommand{\vect}[1]{\ensuremath{\boldsymbol{#1}}}

\newcommand{\smu}{Department of Physics,
Southern Methodist University, 3215 Daniel Ave, Dallas, TX 75275, USA}

\begin{document}

\title{Reconstructing Patchy Reionization with Deep Learning}
\author{Eric~Guzman}
\affiliation{\smu}
\author{Joel~Meyers}
\affiliation{\smu}
\date{\today}

\begin{abstract}
The precision anticipated from next-generation cosmic microwave background (CMB) surveys will create opportunities for characteristically new insights into cosmology.
Secondary anisotropies of the CMB will have an increased importance in forthcoming surveys, due both to the cosmological information they encode and the role they play in obscuring our view of the primary fluctuations.
Quadratic estimators have become the standard tools for reconstructing the fields that distort the primary CMB and produce secondary anisotropies. While  successful for lensing reconstruction with current data, quadratic estimators will be sub-optimal for the reconstruction of lensing and other effects at the expected sensitivity of the upcoming CMB surveys.
In this paper we describe a convolutional neural network, ResUNet-CMB, that is capable of the simultaneous reconstruction of two sources of secondary CMB anisotropies, gravitational lensing and patchy reionization.
We show that the ResUNet-CMB network significantly outperforms the quadratic estimator at low noise levels and is not subject to the lensing-induced bias on the patchy reionization reconstruction that would be present with a straightforward application of the quadratic estimator.
\end{abstract}

\maketitle

\section{Introduction}\label{sec:Intro}
Observations of the cosmic microwave background (CMB) have provided a wealth of cosmological information.  The coming generation of CMB surveys, including those from Simons Observatory~\cite{Ade:2018sbj}, CCAT-prime~\cite{Aravena:2019tye}, CMB-S4~\cite{Abazajian:2016yjj}, PICO~\cite{Hanany:2019lle}, and CMB-HD~\cite{Sehgal:2019ewc}, will map the CMB with unprecedented precision.  In addition to providing greater sensitivity to primary CMB intensity and polarization fluctuations, next generation observations will allow for much more precise measurements of secondary CMB anisotropies~\cite{Aghanim:2007bt}.

Secondary anisotropies of the CMB are generated by interaction of CMB photons with structure which intervenes between our telescopes and the surface of last scattering.
Prominent sources of secondary anisotropies include weak gravitational lensing by large-scale structure (see~\citep{Lewis:2006fu} for a review); the various manifestations of the Sunyaev-Zel'dovich  effect~\citep{Zeldovich:1969ff,Sunyaev:1970er,Sunyaev:1972eq,Sunyaev:1980vz,Sazonov:1999zp}, which describe Compton scattering of CMB photons with free electrons in galaxy clusters and the intergalactic medium; the integrated Sachs-Wolfe~\citep{1967ApJ...147...73S} and Rees-Sciama effects~\citep{1968Natur.217..511R}, characterizing how time-dependent gravitational potentials alter the energy of CMB photons; and the moving lens effect~\citep{1983Natur.302..315B}, where the transverse motion of massive objects imprints temperature fluctuations on the CMB.  The sensitivity of upcoming experiments should allow some of these effects to be detected for the first time in the coming years~\cite{Smith:2016lnt,Louis:2017hoh,Deutsch:2017cja,Deutsch:2017ybc,Meyers:2017rtf,Smith:2018bpn,Hotinli:2018yyc,Yasini:2018rrl,Hotinli:2020ntd}.  These secondary anisotropies act both as a source of cosmological information and as a source of confusion for the primary anisotropies.

Sources of secondary anisotropies can be reconstructed due to the changes they imprint on the statistics of the primary CMB fluctuations. In particular, the primary CMB anisotropies are expected to be well described by nearly Gaussian fluctuations that can be fully characterized by their power spectra $C_\ell$. Sources of secondary anisotropies break rotational invariance, induce off-diagonal correlations between the spherical harmonic coefficients, and cause the observed statistics to be non-stationary.  This change to the statistics can be observed in the data, and thereby can be used to reconstruct the field causing the distortion~\cite{Hu:2001kj,Okamoto:2003zw}.

The standard technique for reconstructing distortions of the CMB fluctuations utilizes a quadratic estimator, constructed from a weighted product of two factors of CMB fluctuations~\cite{Hu:2001kj}.  The quadratic estimator has been used to great success to detect the effects of gravitational lensing at high significance with existing CMB surveys~\cite{Aghanim:2018oex,Wu:2019hek,Darwish:2020fwf}. While the quadratic estimator works well with current data, it will be sub-optimal at the sensitivity anticipated in the next generation of CMB surveys. One reason for this is that higher order effects of the sources of secondary anisotropies become too important to ignore in very high fidelity CMB maps, and use of the quadratic estimator results in an estimate of the distortion field whose variance is limited by secondary anistropies rather than by instrumental noise.  This challenge can be overcome by using maximum likelihood estimators~\cite{Hirata:2002jy,Hirata:2003ka} or by iteratively removing the effects of the secondaries~\cite{Smith:2010gu}. Another limiting factor in the use of the quadratic estimator is that the presence of more than one distortion field can lead to biases in the reconstructed fields.  This bias can be avoided at the cost of increased variance with the use of `bias-hardened' estimators~\cite{Namikawa:2012pe}.

One of the main motivations for an accurate reconstruction of the CMB lensing field is related to the search for primordial gravitational waves.  CMB polarization can be split into curl-free $E$ modes and divergence-free $B$ modes, the latter of which are generated by primordial gravitational waves, but not by density fluctuations at linear order~\cite{Kamionkowski:1996ks,Seljak:1996gy}. Gravitational lensing of the CMB can convert $E$-mode polarization into $B$-mode polarization, and these lensing-induced $B$ modes act as a source of confusion in searches for primordial gravitational waves~\cite{Zaldarriaga:1998ar,Lewis:2001hp}.  With an estimate of the lensing field, it is possible to delens the CMB, thereby removing some of the lensing $B$ modes enabling greater sensitivity to primordial gravitational waves~\cite{Knox:2002pe,Kesden:2002ku,Seljak:2003pn}.

Delensing the CMB is useful beyond the benefits it provides for primordial gravitational wave searches~\cite{Green:2016cjr}.  For example, delensing the temperature and $E$-mode polarization leads to sharper peaks in the power spectra, allowing for a better determination of parameters which impact peak positions~\cite{Baumann:2015rya,Green:2016cjr}.  Delensing can also reduce the off-diagonal covariance induced by lensing~\cite{Green:2016cjr} and thereby improve the sensitivity of the CMB to other effects, such as primordial non-Gaussianity~\cite{Coulton:2019odk}.

In order to achieve the goals of next generation CMB surveys, delensing will be necessary (see e.g.~\cite{Abazajian:2016yjj,Abazajian:2019eic,Abazajian:2020dmr}).  However, achieving the required delensing performance is not without challenges. Delensing algorithms that go beyond the quadratic estimator are under development, but the implementations can be computationally intensive~\cite{Hirata:2003ka,Millea:2017fyd,Millea:2020cpw,Diego-Palazuelos:2020lme}.  In the near-term, delensing the CMB using external tracers of the lensing field holds promise~\cite{Sherwin:2015baa}, though at improved sensitivity, internal delensing, whereby the CMB is delensed using a map of the lensing field reconstructed from the observed CMB map, will be necessary to achieve the best delensing performance~\cite{Smith:2010gu}.  Internal delensing can lead to biases in the delensed CMB map, though these biases can be mitigated without much loss in sensitivity by splitting the data appropriately~\cite{Sehgal:2016eag}.  

Reionization, the process by which the neutral gas filling the Universe became ionized by the first stars and galaxies between redshifts of about 20 and 6, was a complex and non-uniform process~\cite{Barkana:2000fd,Wyithe:2002qu,Barkana:2003qk,Furlanetto:2004nh,McQuinn:2006et,Mesinger:2007pd,Battaglia:2012id,Mitra:2016olz,Dayal:2018hft}. The spatial variations or patchiness in the process of reionization leads to effects on the observed CMB temperature and polarization~\cite{Hu:1999vq,Santos:2003jb,Zahn:2005fn,McQuinn:2005ce,Dore:2007bz,Dvorkin:2008tf,Dvorkin:2009ah,Battaglia:2012im,Park:2013mv,Alvarez:2015xzu,Paul:2020fio,Choudhury:2020kzh}. Patchy reionization can produce $B$-mode polarization, either by the modulation of $E$-mode polarization or by scattering of remote temperature quadrupoles, and these secondary $B$ modes can potentially act as a source of confusion in searches for primordial gravitational waves~\cite{Mukherjee:2019zlb}.

Quadratic estimators have been designed to reconstruct the spatial variations in optical depth that result from patchy reionization~\cite{Dvorkin:2008tf,Dvorkin:2009ah}.  However, when applied to maps of the lensed CMB, the patchy reionization quadratic estimator will be biased by the effects of lensing, since both effects lead to mode-coupling and the estimators are not independent~\cite{Su:2011ff}.  In principle, the lensing estimator is also biased by the presence of patchy reionization, though the bias to the lensing estimate is expected to be small because the effects of patchy reionization are much smaller than those of lensing.  These biases can be mitigated with delensing, and they can be essentially avoided with bias-hardening at the cost of a small increase to the variance of the estimator~\cite{Su:2011ff}.  So far, only upper limits on the effects of patchy reionization exist~\cite{Gluscevic:2012qv,Namikawa:2017uke}, though the effects may be detectable in future data~\cite{Roy:2018gcv}.

In this paper, we apply the techniques of machine learning to reconstruct the effects of patchy reionization in simulated CMB data.  It has been shown recently that machine learning is capable of producing promising results for lensing reconstruction and delensing at noise levels where the quadratic estimator is sub-optimal~\cite{Caldeira:2018ojb}.  The focus of this work is to extend the techniques of Ref.~\cite{Caldeira:2018ojb} to the simultaneous reconstruction of CMB lensing and patchy reionization.  As described above, reconstructing patchy reionization in the presence of lensing is challenging due to the extra variance coming from lensing-induced $B$-mode polarization and by the bias that can result from the sensitivity of the patchy reionization estimator to the effects of lensing.  A main goal of this work is to determine the extent to which machine learning is capable of surmounting these challenges.

\section{Quadratic Estimator}\label{sec:QE}
In this section we will review the construction of the quadratic estimator for CMB lensing~\cite{Hu:2001kj} and for patchy reionization~\cite{Dvorkin:2008tf}.  The quadratic estimator has become the standard technique for lensing reconstruction with current data, though it will be sub-optimal with the high precision data expected from future CMB surveys~\cite{Hirata:2003ka}.  We will also discuss how the estimator for the spatial variation of the optical depth is biased in the presence of lensing~\cite{Su:2011ff}.

\subsection{Weak Gravitational Lensing}\label{subsec:QE_lensing}

Cosmological structure which intervenes between our telescopes and the CMB surface of last scattering generates gravitational potentials which deflect CMB photons~\cite{Lewis:2006fu}. In real space this effect can be understood as a re-mapping of the unlensed CMB by a direction-dependent deflection angle, 
\begin{align}
    \label{eqn:LensedFields}
    \lenmap{T}(\nhat) &= \unlmap{T}(\nhat + \nabla \phi (\nhat)) \, , \nonumber\\
    (\lenmap{Q} \pm i\lenmap{U})(\nhat) &= (\unlmap{Q} \pm i\unlmap{U})(\nhat + \nabla \phi (\nhat)) \, ,
\end{align}
where $(\unlmap{T},\unlmap{Q},\unlmap{U})$ are the primordial CMB fields, $\phi$ is the lensing potential, $\nhat$ is the line-of-sight direction, and the superscript `len' refers to the lensed field~\cite{Hu:2001kj}. In harmonic space, lensing of the CMB induces mode-coupling, or correlation, between fluctuations of different wavenumbers $\ell_1 \neq \ell_2$ that is proportional to the lensing potential $\phi$. In the flat-sky approximation~\cite{Hu:2001kj},
\begin{equation}
    \label{eqn:initQE}
    \left\langle \lenmap{X}(\vect{\ell}_{1}) \lenmap{Y}(\vect{\ell}_{2})  \right\rangle_{\mathrm{CMB}} \propto f^{\phi}_{XY}(\vect{\ell}_{1},\vect{\ell}_{2}) \phi(\vect{\ell}) \, ,
\end{equation}
where $\vect{\ell} = \vect{\ell}_{1} + \vect{\ell}_{2}$ and $(X,Y)=(T,E,B)$.
In this paper we focus on the $EB$ mode-coupling since the corresponding estimator has the lowest variance for the noise levels we consider. For this case, the factor $f^{\phi}_{XY}(\vect{\ell}_{1},\vect{\ell}_{2})$ in Eq.~\eqref{eqn:initQE} is defined as
\begin{align}
    \label{eqn:f_phi}
    f^{\phi}_{E B}(\vect{\ell}_{1}, \vect{\ell}_{2}) &= \left[(\vect{\ell \cdot \ell_{1}})C_{\ell_{1}}^{E E} - (\vect{\ell \cdot \ell_{2}})C_{\ell_{2}}^{B B}\right] \nonumber \\
    &\qquad \qquad \qquad \times \sin{2(\varphi(\vect{\ell}_1) - \varphi(\vect{\ell}_2))} \, .
\end{align}

The quadratic estimator provides a way of estimating any field leading to mode coupling by using a weighted product of pairs of observed CMB maps. The unbiased minimum-variance $EB$ quadratic estimator for the CMB gravitational lensing potential is given by~\cite{Hu:2001kj}
\begin{equation}
    \label{eqn:LensingRecon}
    \hat{\phi}_{E B}(\vect{\ell}) = N^{\phi}_{E B}(\vect{\ell}) \int{\frac{d^2 \vect{\ell}_{1}}{(2 \pi)^2 } \obsmap{E}(\vect{\ell}_{1}) \obsmap{B}(\vect{\ell}_{2}) F^{\phi}_{E B}(\vect{\ell}_{1} \, , \vect{\ell}_{2})} \, ,
\end{equation}
with
\begin{equation}
    F^{\phi}_{E B}(\vect{\ell}_{1}, \vect{\ell}_{2}) = \frac{f^{\phi}_{E B}(\vect{\ell}_{1}, \vect{\ell}_{2})}{C_{\ell_{1}}^{E E,\mathrm{obs}} C_{\ell_{2}}^{B B,\mathrm{obs}}} \, .
\end{equation}
The factor in front of Eq.~\eqref{eqn:LensingRecon} is an overall normalization factor defined as
\begin{equation}
    \label{eqn:LensingReconNoisePower}
    N^{\phi}_{E B}(\vect{\ell}) = \left[\int{\frac{d^2 \vect{\ell}_{1}}{(2 \pi)^2 } f^{\phi}_{E B}(\vect{\ell}_{1}, \vect{\ell}_{2}) F^{\phi}_{E B}(\vect{\ell}_{1}, \vect{\ell}_{2})}\right]^{-1} \, ,
\end{equation}
which also gives the variance of the estimator.
We define each observed field as the lensed field plus noise, 
\begin{equation}
    \label{eqn:LensObservedMapDef}
    \obsmap{X}(\vect{\ell}) = \lenmap{X}(\vect{\ell}) + N^{X}(\vect{\ell}) \, ,
\end{equation}
where $X=(T,E,B)$ and $N^{X}(\vect{\ell})$ is the noise corresponding to field $X$, which we take to be a Gaussian random field whose power spectrum is given by
\begin{align}
    C^{T T, \mathrm{noise}}_{\ell} &= \Delta_{T}^{2} e^{\ell^{2} \theta_{\mathrm{FWHM}}^{2} / (8 \ln{2})} \, ,\nonumber\\
    C^{E E, \mathrm{noise}}_{\ell} &= C^{B B, \mathrm{noise}}_{\ell} = \Delta_{P}^{2} e^{\ell^{2} \theta_{\mathrm{FWHM}}^{2} / (8 \ln{2})} \, .
\end{align}

At the noise levels of future experiments, the lensing quadratic estimator is sub-optimal, and maximum likelihood techniques or an iterative delensing procedure is necessary to produce the best estimate of the lensing field~\cite{Hirata:2003ka,Smith:2010gu}.  The basic idea of the iterative technique is to first apply the quadratic estimator on the observed maps, then use the resulting estimate of the lensing field to delens, next apply the quadratic estimator on the delensed maps, delens again with the resulting estimate of the residual lensing field, and iterate this procedure until convergence is achieved~\cite{Smith:2010gu}.  Simulations have shown that the iterative technique closely matches the performance of the maximum likelihood estimator~\cite{Hirata:2003ka,Smith:2010gu}.

\subsection{Patchy Reionization}\label{subsec:QE_tau}

The inhomogeneous nature of the reionization epoch causes several observable impacts on the CMB~\cite{Hu:1999vq,Santos:2003jb,Zahn:2005fn,McQuinn:2005ce,Dore:2007bz,Dvorkin:2008tf,Dvorkin:2009ah,Battaglia:2012im,Park:2013mv,Alvarez:2015xzu}.  
The patchiness of reionization leads to an anisotropic optical depth whose effects on the CMB can be separated into three categories: screening, scattering, and the kinetic Sunyaev-Zel'dovich effect.

First, the variation of the optical depth on the sky, $\tau(\nhat)$, leads to a spatially dependent screening of CMB fluctuations due to the scattering of CMB photons into and out of our line of sight.  Next, Thomson scattering of remote temperature quadrupoles on the free electrons in ionized bubbles generates new polarization fluctuations.
Finally, the radial velocity of ionized bubbles generates CMB temperature fluctuations through the kinetic Sunyaev-Zel'dovich effect. 


For the purpose of this paper, we focus solely on the screening effect of patchy reionization, since we are interested in the effect that can be reconstructed from observations of CMB polarization alone.  The scattering effect results in new polarization that is not correlated with the polarization generated at the surface of last scattering, and the kinetic Sunyaev-Zel'dovich effect only produces temperature fluctuations.

The screening effect modulates the amplitude of the temperature and polarization anisotropies, scaling them by a factor $e^{-\tau(\nhat)}$. For the purpose of the simulations described below, we first apply modulation followed by lensing giving maps in real space according to
\begin{align}
    \label{eqn:ModulatedFields}
    \modmap{T}(\nhat) &= \lenmap{( \unlmap{T}(\nhat) e^{-\tau(\nhat)} )} \nonumber\\
    (\modmap{Q} \pm i\modmap{U})(\nhat)  &= \lenmap{( (\unlmap{Q} \pm i\unlmap{U})(\nhat) e^{-\tau(\nhat)} )}\, ,
\end{align}
such that the superscript 'mod' refers to maps that have been first  modulated by patchy reionization then lensed. This is an idealized treatment of the two effects, which in reality cannot be so cleanly separated in time.
We further assume for our simulations that the lensing and patchy reionization fields are uncorrelated, though in reality they should exhibit some correlation, since they develop from the same underlying density fluctuations.

Following the formalism for the lensing quadratic estimator, an unbiased minimum variance $EB$ estimator can be derived for $\tau(\nhat)$ (see \cite{Dvorkin:2008tf, Su:2011ff} for the full derivation) which takes the form
\begin{equation}
    \label{eqn:PatchyTauRecon}
    \hat{\tau}_{E B}(\vect{\ell}) = N^{\tau}_{E B}(\vect{\ell}) \int{\frac{d^2 \vect{\ell}_{1}}{(2 \pi)^2 } \obsmap{E}(\vect{\ell}_{1}) \obsmap{B}(\vect{\ell}_{2}) F^{\tau}_{E B}(\vect{\ell}_{1} \, , \vect{\ell}_{2})} \, ,
\end{equation}
where like Eq.~\eqref{eqn:LensingReconNoisePower}, the factor 
\begin{equation}\label{eqn:Nltau_QE}
    N^{\tau}_{E B}(\vect{\ell}) = \left[\int{\frac{d^2 \vect{\ell}_{1}}{(2 \pi)^2 } f^{\tau}_{E B}(\vect{\ell}_{1}, \vect{\ell}_{2}) F^{\tau}_{E B}(\vect{\ell}_{1}, \vect{\ell}_{2})}\right]^{-1} \, ,
\end{equation}
is a normalization that is equal to the variance of the estimator, with the mode coupling caused by patchy screening given by
\begin{equation}
    f^{\tau}_{E B}(\vect{\ell}_{1}, \vect{\ell}_{2}) = \left[C_{\ell_{1}}^{E E} - C_{\ell_{2}}^{B B}\right] \sin{2(\varphi(\vect{\ell}_1) - \varphi(\vect{\ell}_2))} \, ,
\end{equation}
and the filter is
\begin{equation}
    F^{\tau}_{E B}(\vect{\ell}_{1}, \vect{\ell}_{2}) = \frac{f^{\tau}_{E B}(\vect{\ell}_{1}, \vect{\ell}_{2})}{C_{\ell_{1}}^{E E,\mathrm{obs}} C_{\ell_{2}}^{B B,\mathrm{obs}}} \, .
\end{equation}
In Eq.~\eqref{eqn:PatchyTauRecon} and throughout the rest of the paper, the $\obsmap{E},\obsmap{B}$ are defined according to
\begin{equation}
    \label{eqn:FinObservedMapDef}
    \obsmap{X}(\vect{\ell}) = \modmap{X}(\vect{\ell}) + N^{X}(\vect{\ell}) \, ,
\end{equation}
with $X$ and $N^{X}(\vect{\ell})$ as described below Eq.~\eqref{eqn:LensObservedMapDef}.

In the presence of lensing, the $\tau$ estimator is biased by a spurious signal that results from the mode-coupling induced by lensing. The amplitude of the bias is more than an order of magnitude greater than the true patchy reionization signal, and the bias remains significant even after 98\% of the lensing signal has been removed~\cite{Su:2011ff}. The bias for the $EB$ estimator can be calculated by how much of the lensing mode-coupling is picked up by the patchy reionization estimator~\cite{Su:2011ff}
\begin{equation}
    \label{eqn:LensingBiasTh}
    \beta_{EB}(\vect{\ell}) = N^\tau_{E B}(\vect{\ell}) \int{\frac{d^2 \vect{\ell}_{1}}{(2 \pi)^2 } f^{\phi}_{E B}(\vect{\ell}_{1}, \vect{\ell}_{2}) F^{\tau}_{E B}(\vect{\ell}_{1}, \vect{\ell}_{2}) \phi(\vect{\ell})} \, .
\end{equation}
It is possible to construct a bias-hardened estimator that is insensitive to the effects of lensing and which does not have significantly higher variance than the estimator described above~\cite{Su:2011ff}.

In addition to the bias, lensing also adds variance to the patchy reionization estimator by increasing the observed $B$-mode polarization power. A strategy to reconstruct the screening effect of patchy reionization using the tools above is therefore to first reconstruct the lensing field, use that estimated lensing field to delens the CMB, and finally to reconstruct the patchy reionization using a bias-hardened estimator. This procedure is difficult to implement in practice and would lead to complicated noise properties in the resulting estimates.
One of the main goals of this paper is to compare the performance of an idealized version of the above procedure to the results that can be obtained through machine learning. We will show how our network is capable of making simultaneous estimates of a delensed polarization map, a lensing map, and a patchy reionization map in a single step.

\section{Deep Learning Network and Methods}\label{sec:DeepLearning}

In this section, we provide a brief introduction to machine learning and its applications to image processing. We then discuss the design of the ResUNet-CMB architecture that we apply to the task of patchy reionization reconstruction with simulated CMB maps.

\subsection{Deep Neural Networks}
\begin{figure*}[ht!]
    \centering
    \includegraphics[width=0.95\textwidth]{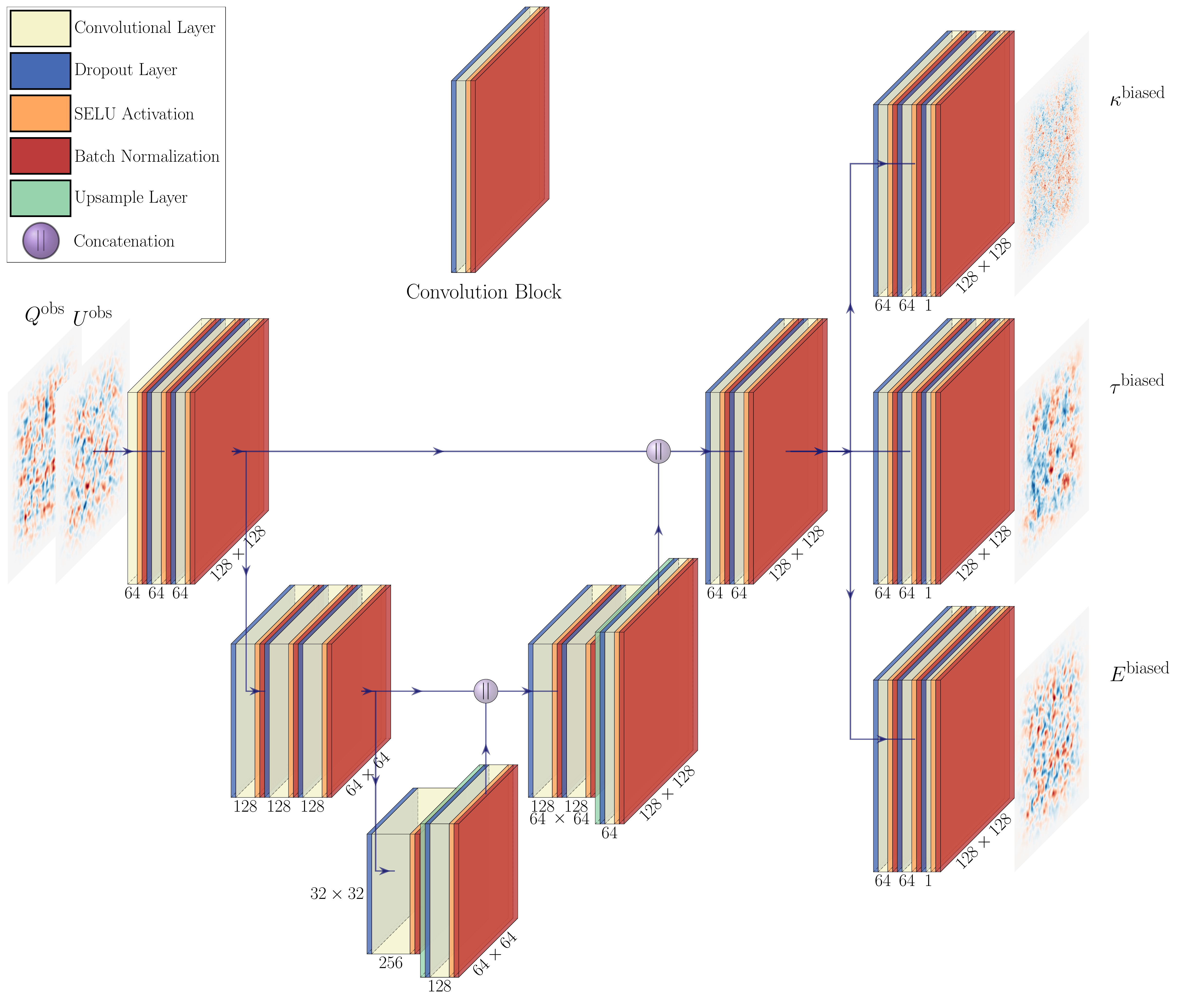}
    \caption{ResUNet-CMB architecture with residual connections excluded for clarity\protect\footnotemark[1]. The input images, ($\obsmap{Q}$,$\obsmap{U}$) are concatenated along the channel dimension.
    Convolutional layers are labeled with the number of filters they contain. The final images, ($\biasmap{\kappa}$,$\biasmap{\tau}$,$\biasmap{E}$), are the output of the final batch normalization layer of each branch. The network has 5,292,367 total parameters and 5,287,113 trainable parameters. The receptive field size when ignoring residual and skip connections
    is $101\times101$ pixels for each output. Widths of layers are chosen for visual clarity and are not to scale.}
    \label{fig:ResUNet-Structure}
\end{figure*}

Machine learning is a process whereby computers learn the rules to complete tasks when provided a set of examples. In the past decade, artificial neural networks (ANNs) have become ubiquitous in machine learning, being applied to a wide range of problems. The method is inspired by biological systems in that ANNs use neurons as the building blocks. These neurons are grouped into layers and carry out tensor operations.

Deep learning makes use of these neural networks by using multiple successive layers of neurons, offering the ability to learn complex nonlinear relationships hidden in sets of data. There is a significant computational cost in training deep learning networks, but once trained, the networks are capable of very quickly making predictions with very low resource requirements. When dealing with image data in computer vision tasks, the convolutional neural network (CNN) has become the dominant deep learning network design.

It is natural to ask whether deep learning can be usefully applied to cosmology.
Maps of the content of the Universe (such as the CMB) are 2D or 3D images, from which we can extract cosmological information using CNNs. As a result, deep learning applications have a growing presence in cosmology. Some recent applications of deep learning to the CMB involving CNNs are full-sky foreground removal \cite{Petroff:2020fbf}, fast Wiener filtering \cite{Munchmeyer:2019kng}, mass estimation of galaxy clusters \cite{Gupta:2020him,Gupta:2020yvd}, and cosmic string detection \cite{Ciuca:2017gca}.

Cosmological structure along our line of sight leaves characteristic distortions imprinted in the observed CMB polarization maps resulting in secondary anisotropies which can be identified and extracted by a CNN. These distortions are correlated over a range of scales, and so the hierarchical feature extraction of the CNN is well-suited to detect the impact of these distortions.
In this paper, we employ a particular type of CNN called a ResUNet to analyze secondary anisotropies of the CMB.  This choice is motivated by the success of a ResUNet for flat-sky CMB lensing reconstruction~\cite{Caldeira:2018ojb}.
In the sections that follow we discuss the ingredients that define a ResUNet, we describe the specific ResUNet architecture we designed for simultaneous lensing and patchy reionization reconstruction, and we specify how our network is implemented and trained.

\subsection{Residual U-Network (ResUNet)} %

Convolutional layers are the core of a CNN. Each neuron in a convolutional layer operates on a region of pixels of its input called the receptive field. In the first convolutional layer of a network, the receptive field is completely determined by the kernel size. The kernel size specifies the dimensions of the $k\times k$ filters which are slid across the input to extract features. Each filter is sensitive to a unique feature. Common choices for the kernel size use $k=3,5,7,...,$ with odd numbers selected so symmetry about the central pixel can be maintained to reduce edge effects. For more details about convolutional layers see Ref.~\cite{dumoulin2018guide}.

By chaining together convolutional layers, neurons in each successive layer become connected to a larger region of the initial input layer thereby increasing the size of each consecutive receptive field. The progressive increase in receptive field size of each convolutional layer makes the network sensitive to high-level features of larger scales. For tasks involving images, using a CNN is beneficial since mapping a neuron directly to every pixel would be computationally intractable.

Convolutional autoencoders are a type of CNN that take a tensor for input and produce an identically sized one for output. When used with image data, the tensor takes a shape of (image height, image width, number of channels). For example, if we have a $128\times128$ pixel RGB image, then the input tensor shape would be (128, 128, 3) where the red, green, and blue pixel values each represent a different channel. Employed with supervised learning, convolutional autoencoders are able to learn a nonlinear mapping between the input and output space.

There are two primary components in a convolutional autoencoder, the encoder and decoder. In the encoding phase, relevant features for the reconstruction of the output are learned by each convolutional layer and increase in complexity, becoming more abstract in deeper layers of the network. The output of each convolution is called a feature map since it is a tensor made of encoded features. The feature maps are continuously down-sampled at regular intervals lowering their dimensionality until a set minimum size is reached. This down-sampling helps increase the receptive field with respect to the input image while also reducing the propagation of unnecessary information.
\footnotetext{Graphic made with publicly available code from \url{https://github.com/HarisIqbal88/PlotNeuralNet}.}
%
%

The decoder takes the features learned in the last layer of the encoder, the minimum representation, and learns how to combine them to achieve the desired output. Starting from the minimum representation which is an abstract base, it is up-sampled while detail is added through consecutive convolutional layers. The final result is an output image with the same height and width as the initial input to the network.

Some localization information is lost through down-sampling. To combat this, the outputs of some layers in the encoder phase can be concatenated with those of the same size in the decoder. The concatenations are with respect to the channel dimension and are called skip connections. As the output is reconstructed from the minimum representation, the skip connections provide high resolution information just after up-sampling. These skip connections in combination with a convolutional autoencoder were introduced for the purpose of biomedical segmentation, and the resulting architecture is called a U-Net \cite{ronneberger2015unet}. The encoding phase and decoding phase in the U-Net often mirror each other in both depth and the locations at which up- or down-sampling occur. As a result, this type of network is often presented in a U-shape, hence its name.


As the network grows deeper by stacking convolutional layers, it has been shown that performance and accuracy of the network decreases~\cite{he2014convolutional,he2015deep,srivastava2015highway}.
Residual connections, which take the input of a convolutional layer and add it element-wise to the output of a subsequent convolutional layer, allow more layers to be added to the network without degrading performance by helping to propagate parameter corrections to earlier layers and information to latter ones, through direct identity paths~\cite{he2015deep,he2016identity,balduzzi2018shattered}. Their use can help improve network training and results. Adding residual and skip connections to a convolutional autoencoder yields an architecture called a ResUNet~\cite{KayalibayJS17, MilletariNA16, Zhang_2018}.


\subsection{Network Architecture (ResUNet-CMB)}
We focus here on the application of a ResUNet designed to reconstruct the effects of patchy reionization on the CMB polarization in the presence of lensing. After training, the ResUNet takes as input maps of the observed $Q$ and $U$ polarization $(\obsmap{Q}, \obsmap{U})$ and produces map-level estimates of the lensing convergence, the patchy reionization, and the primordial $E$-mode polarization $({\kappa},{\tau},\unlmap{E})$.

We use a modified version of the architecture defined in Ref.~\cite{Caldeira:2018ojb} which was shown to work in reconstructing the lensing convergence map and therefore provided an ideal starting point from which to build upon. Implementation of our network, which we call ResUNet-CMB, was done with the Keras package of TensorFlow 2.0\footnote{The code for ResUNet-CMB and the data pipeline used can be found at \url{https://github.com/EEmGuzman/resunet-cmb}.}.

A schematic representation of the ResUNet-CMB architecture is shown in Fig.~\ref{fig:ResUNet-Structure} where the residual connections were left out for clarity. The network is made up of a series of convolution blocks. 
We define a convolution block as consisting of four components in the following order: dropout layer, convolutional layer, activation layer, and batch normalization layer. The convolutional layer is set with a kernel size of $5\times5$ and 'same' padding. For the activation we use the Scaled Exponential Linear Unit (SELU) function~\cite{klambauer2017selfnormalizing}. The dropout layer, with a dropout rate of 0.3 which we found to be the optimal value, is necessary to prevent the network from over-fitting. The dropout layer is omitted from the first convolution block in the network in order to prevent the irretrievable loss of initial information.

The final block of each branch in the decoder phase differs from the others in that the activation function is switched to a linear one. Down-sampling in the encoder phase is achieved by setting the stride to 2 in the convolutional layer. The stride refers to the number of pixels between applications of the filter to the input image. A dedicated up-sampling layer with nearest-neighbor interpolation was used for the decoding phase and was placed before the dropout layer to reduce the production of visual artifacts \cite{odena2016deconvolution}.

Skip connections occur every three convolution blocks in the encoder and concatenate along the channel dimension. Residual connections connect every two convolution blocks.   The first residual connection takes the $(\obsmap{Q},\obsmap{U})$ input layer and adds it element-wise to the batch normalization output of the second block. The next connection is from the output of the second block to that of the fourth block. The connections continue this pattern until the final two blocks of the decoder where no residual connections are used.

In both the encoding and decoding phases, some residual connections occur between convolution blocks with different input and output tensor dimensions. For such residual connections, a convolutional layer with linear activation and appropriate stride and number of filters is inserted into the connection in order to reconcile the tensor dimensions.
If there is a difference in the image size between blocks in the decoder phase, a dedicated upsampling layer is used in addition to the convolutional layer.
We found adding a batch normalization layer in every residual connection improved our results by lowering validation loss for all three outputs. In residual connections containing a convolutional layer, the batch normalization is placed after the convolutional layer.

\subsection{Data Pipeline}
\label{sec:DataPipe}

The publicly available \texttt{CAMB}\footnote{\url{https://camb.info}} software is used to produce the primordial CMB and lensing power spectra~\cite{Lewis:1999bs}. A cosmology with parameters of $H_{0} = 67.9$~km~s$^{-1}$~Mpc$^{-1}$, $\Omega_{b}h^{2} = 0.0222$, $\Omega_{c}h^{2} = 0.118$, $n_{s} = 0.962$, $\tau = 0.0943$, and $A_{s} = 2.21\times 10^{-9}$ was chosen since it allows a comparison with results from Ref.~\cite{Caldeira:2018ojb}. We take the patchy reionization spectrum $C_\ell^{\tau\tau}$ to match the $\bar{\tau} = 0.058$ model of Ref.~\cite{Roy:2018gcv}.

Using the theory power spectra from \texttt{CAMB} and a modified version of \texttt{Orphics}\footnote{\url{https://github.com/msyriac/orphics}}, seven different maps are produced as Gaussian random fields: lensing convergence $\kappa\equiv\frac{1}{2}\nabla^2\phi$, patchy reionization $\tau$, primordial $E$-mode polarization $\unlmap{E}$, primordial $Q$ and $U$ polarization $\unlmap{Q}$ and $\unlmap{U}$, and observed $Q$ and $U$ polarization $\obsmap{Q}$ and $\obsmap{U}$. To obtain the observed polarization maps we use Eq.~\eqref{eqn:FinObservedMapDef}, where the primordial maps $(\unlmap{Q}, ~\unlmap{U})$ are first modulated by $\tau$, then lensed with $\kappa$, and finally summed with a noise map as in Eq.~\eqref{eqn:ModulatedFields}. 

Each map is $128 \times 128$ pixels and covers a $5^{\circ} \times 5^{\circ}$ patch of sky.  A cosine taper of $1.5^{\circ}$ was applied to all maps to reduce edge effects. Instrument noise was implemented at the levels of $\Delta_T=0.0~\mu$K-arcmin, 0.2~$\mu$K-arcmin, 1~$\mu$K-arcmin, and 2~$\mu$K-arcmin assuming $\Delta_{P} = \sqrt{2} \Delta_{T}$. We used $\theta_\mathrm{FWHM} = 1.4'$ for the cases of 1~$\mu$K-arcmin and 2~$\mu$K-arcmin, but $\theta_\mathrm{FWHM} = 1.0'$ for the 0.2~$\mu$K-arcmin case, the latter of which was chosen to make direct comparison with the results of Ref.~\cite{Su:2011ff}. 

For each noise level, 70000 sets of seven maps were produced with the random seed set to 1225 with the python \texttt{NumPy} package. In approximately 20\% of the maps selected at random, we set both $\kappa$ and $\tau$ to zero.  We found that including these `null maps' (meaning no lensing or modulation applied to $(\unlmap{Q},\unlmap{U})$) in training helps the network generalize and prevents it from predicting the presence of a signal when none is present in test data. 
In addition to the set of 70000, we also produce a separate collection of 7000 sets of maps, which we call the prediction set, on which most analysis of ResUNet-CMB predictions was conducted. No null maps were included in the prediction set. 

The 70000 maps were split in a 80:10:10 ratio for training, validation, and test sets. Similar to what was done in Ref.~\cite{Chardin:2019euc}, all maps in the training, validation and test sets are standardized with respect to values from the training set according to 
\begin{equation}
\label{eqn:Prep_standard}
    X_\mathrm{Processed} = \frac{\left( X - \overline{X}_\mathrm{Training}\right)}{\sigma_\mathrm{Training}} \, .
\end{equation}
Here $X$ represents a single unprocessed map from any data set. 
The mean and standard deviation are calculated for each map type over the full training set.

ResUNet-CMB predictions are made by feeding in $(\obsmap{Q},\obsmap{U})$ from the prediction set as input to the fully trained network. The input maps are also standardized according to Eq.~\eqref{eqn:Prep_standard} with the mean and standard deviation from the prediction set. Making a single prediction of $({\kappa},{\tau},\unlmap{E})$ regardless of noise level, takes approximately 0.021 seconds on average.

The ResUNet-CMB predictions are post-processed to rescale the outputs to physically meaningful values. This must be done because the network is trained on standardized maps, so the predictions made on the fully trained network will also be standardized. To perform the rescaling, we rearrange Eq.~\eqref{eqn:Prep_standard} to solve for $X$ using the mean and standard deviation from the training set (since with real data we would not have a priori knowledge of the true $({\kappa},{\tau},\unlmap{E})$ maps).

\subsection{Training}
For each instrument noise level a separate network was trained.
We chose to use a batch size of 32 to train the networks.
The initial learning rate was set to 0.25 and decayed by a factor of $0.5$ if after three epochs there was no improvement in validation loss. Individual learning rate decay of each trainable parameter was achieved by using the Adam optimizer with the TensorFlow default parameters~\cite{kingma2014adam}. If there was no improvement in the validation loss after 10 consecutive epochs then training was stopped and the best network saved. Training the noiseless network using a training split of 56000 sets of maps took an average of 10 hours on a single Nvidia Tesla V100 32GB.

\section{Results}

\begin{figure*}[ht!]
    \centering
    \includegraphics[width=\textwidth]{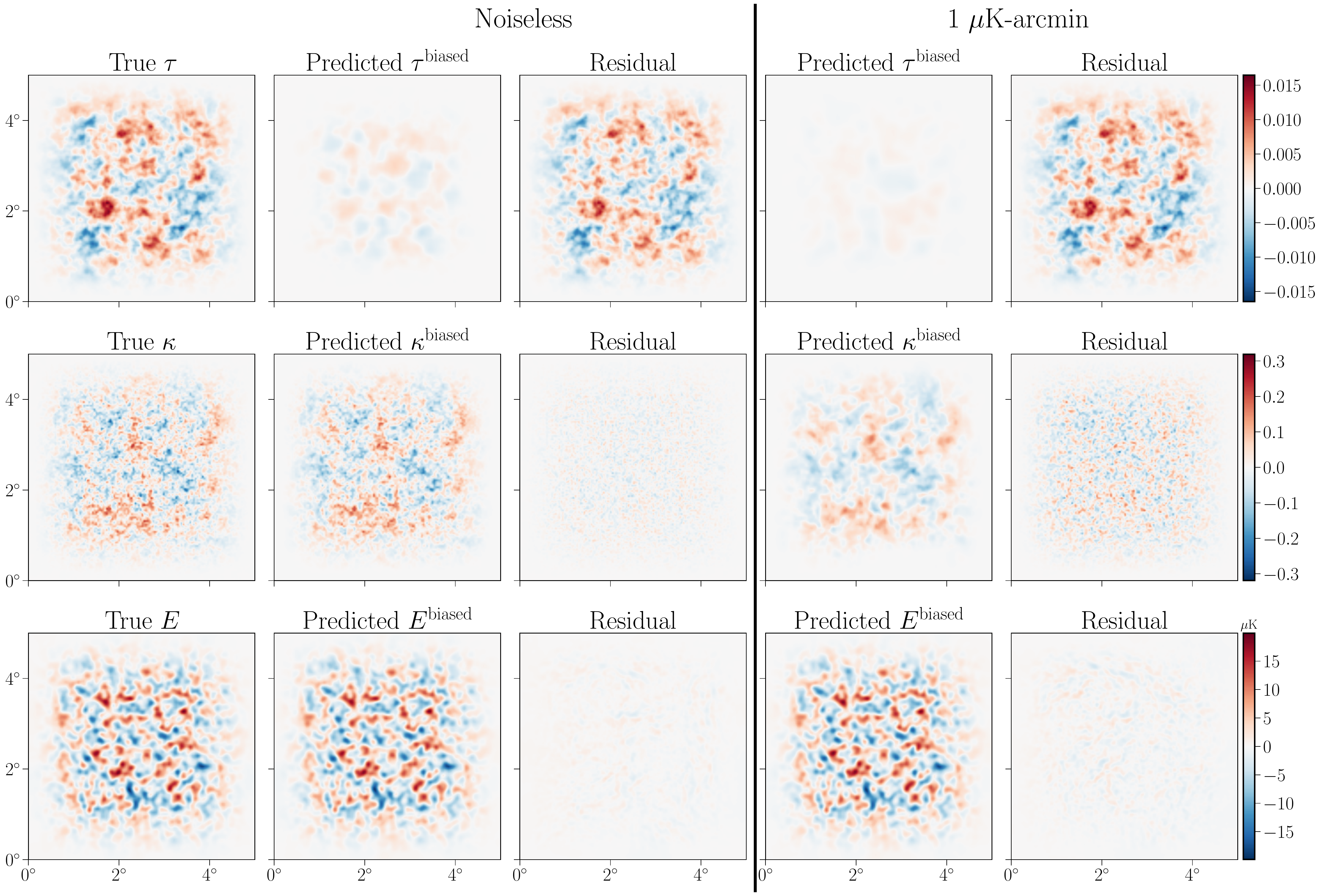}
    \caption{Sample ResUNet-CMB predictions from fully trained networks for two noise levels compared to the true maps. 
    The predicted maps shown here have not been rescaled as in Eq.~\eqref{eqn:norm_prediction} and thus have a multiplicative bias.  Residual maps are calculated as the true map minus the predicted map.}
    \label{fig:map_results_2noiselvls}
\end{figure*}


In this section we examine the performance of the ResUNet-CMB network.  We analyze maps, power spectra, and reconstruction noise curves as evaluation metrics. As described in Section~\ref{sec:DataPipe}, 
we feed in the inputs $(\obsmap{Q},\obsmap{U})$ from the prediction data set to the fully trained network which produces output maps $({\kappa},{\tau},\unlmap{E})$.  The output maps are biased, and so we refer to the direct outputs of the network as $(\biasmap{\kappa},\biasmap{\tau},\biasmap{E})$; we address how to treat the bias later in this section.

\subsection{Maps}
Figure~\ref{fig:map_results_2noiselvls} shows a single prediction for each of $(\biasmap{\kappa},\biasmap{\tau},\biasmap{E})$ from the fully trained ResUNet-CMB network for the noiseless and 1~$\mu$K-arcmin case.  We highlight these results since they allow for partial comparison to results from Ref.~\cite{Caldeira:2018ojb}. Our ResUNet-CMB network was able to reconstruct some of the largest scale features of $\tau(\nhat)$ despite the fact that the impact of patchy reionization on the CMB polarization is orders of magnitude smaller than lensing for the chosen cosmological parameters. The noiseless $\biasmap{\kappa}$ and $\biasmap{E}$ are lacking features on only the smallest scales as can be seen in the residual maps which show the difference between the predicted and truth maps.

With an increase in noise, all predictions degrade with $\biasmap{\kappa}$ and $\biasmap{\tau}$ seeing the largest change. The primary $E$-mode polarization is still faithfully reconstructed on most scales despite increasing instrument noise.
The 1~$\mu$K-arcmin reconstruction of $\biasmap{\tau}$ has residual power that is about 18\% higher for $\ell<840$ than that of the noiseless reconstruction of $\biasmap{\tau}$, where the residual power is defined as the power spectrum of the true map minus the predicted map. There is a larger fractional increase in the residual power for $\biasmap{\kappa}$ with the addition of noise, especially in the range $56<\ell<1500$.
The predictions of $\biasmap{E}$ are most affected by noise on small scales, and the increase in the residual power peaks at about $\ell=2900$ when comparing the noiseless and 1~$\mu$K-arcmin predictions.

\begin{figure}[t]
    \centering
    \includegraphics[width=\linewidth]{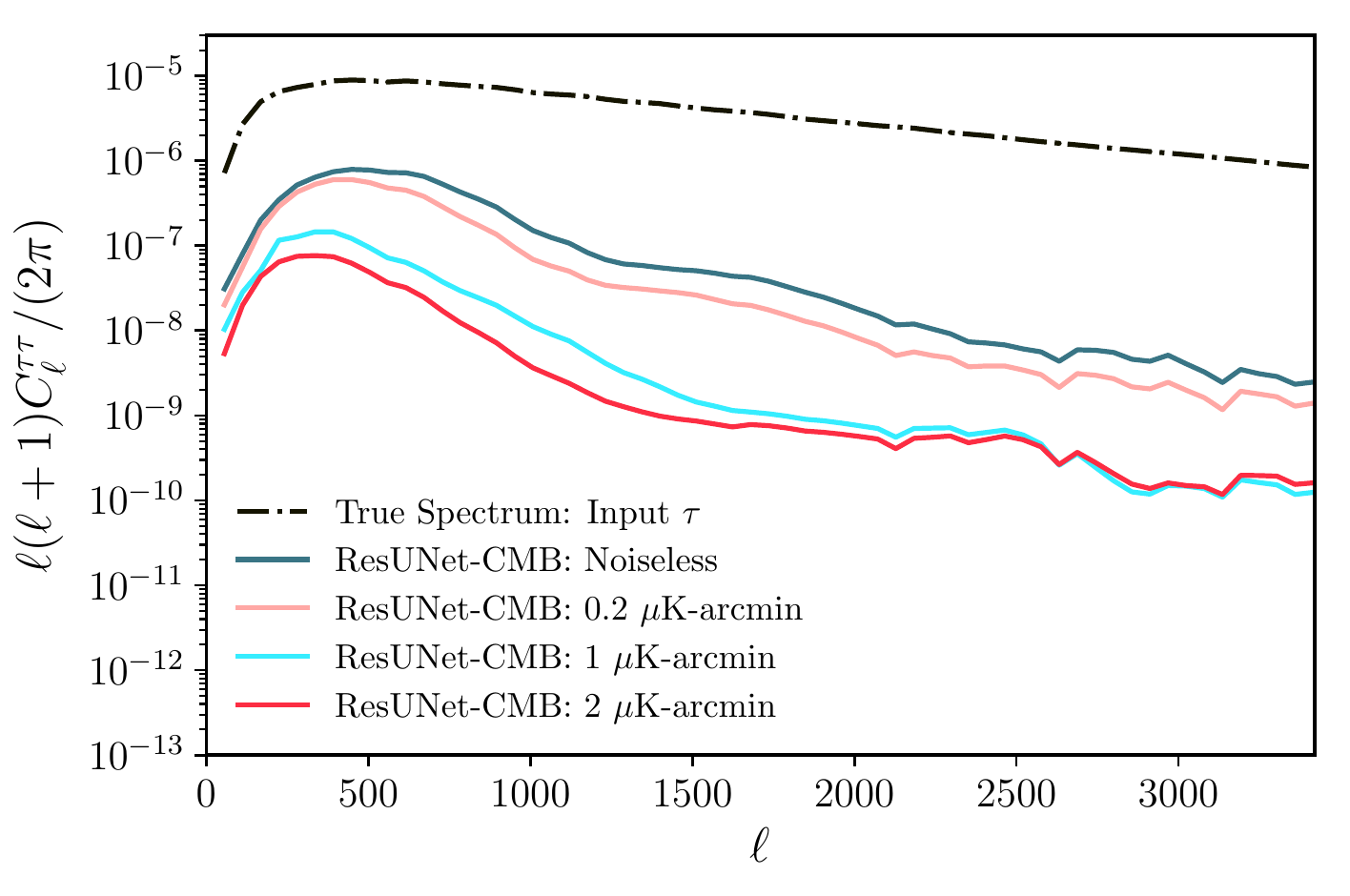}
    \caption{Power spectra of the optical depth fluctuation maps $\biasmap{\tau}$ predicted by ResUNet-CMB  for each of four noise levels averaged over the 7000 realizations included in the prediction set, compared to the averaged power spectrum of the true patchy reionization maps (black dashed-dot).
    }
    \label{fig:tau_ps}
\end{figure}

\begin{figure}[t]
    \centering
    \includegraphics[width=\linewidth]{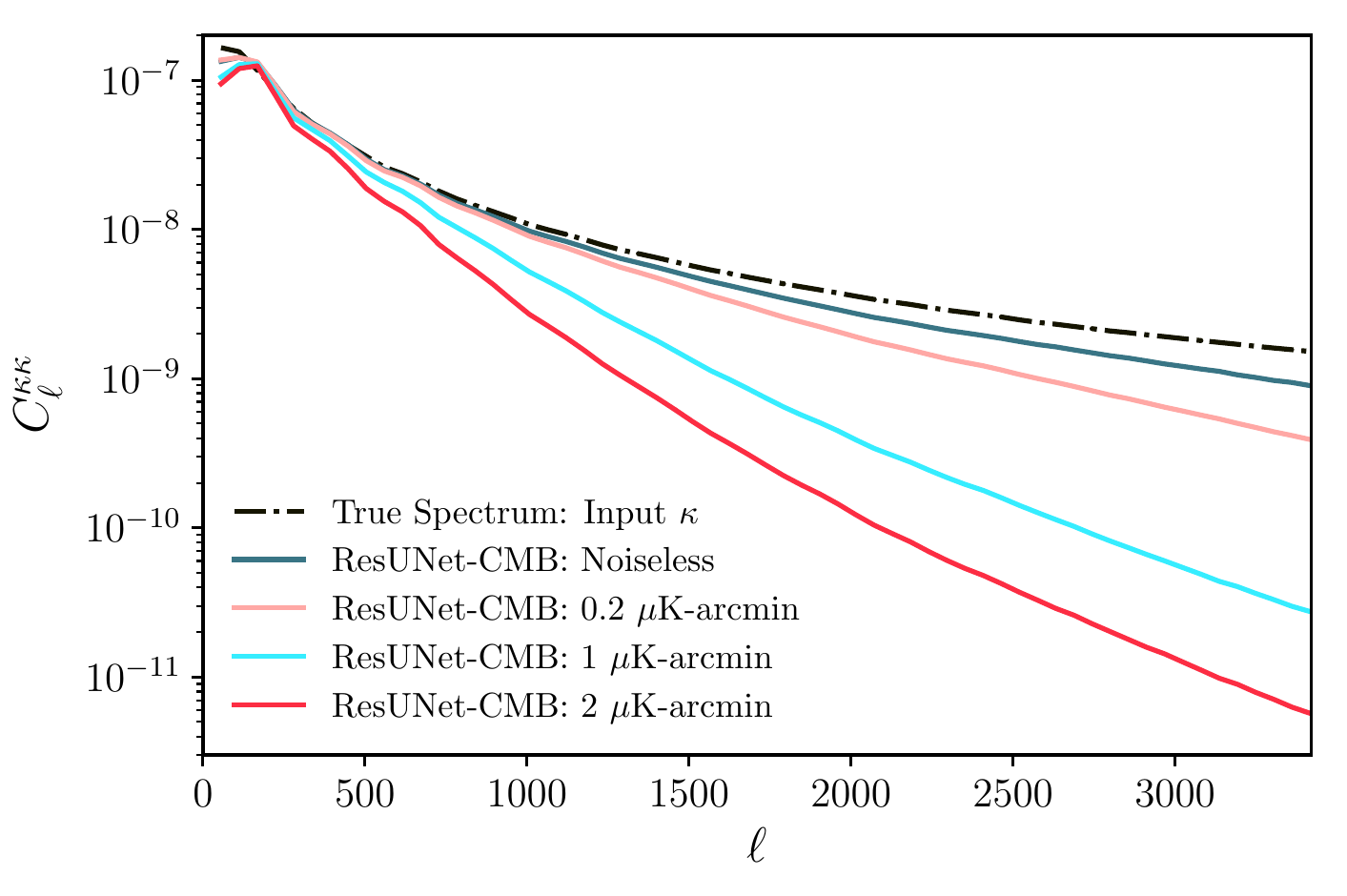}
    \caption{Power spectra of the lensing convergence maps $\biasmap{\kappa}$ predicted by ResUNet-CMB  for each of four noise levels averaged over the 7000 realizations included in the prediction set, compared to the averaged power spectrum of the true lensing convergence maps (black dashed-dot).
    }
    \label{fig:kappa_ps}
\end{figure}

\begin{figure}[t]
    \centering
    \includegraphics[width=\linewidth]{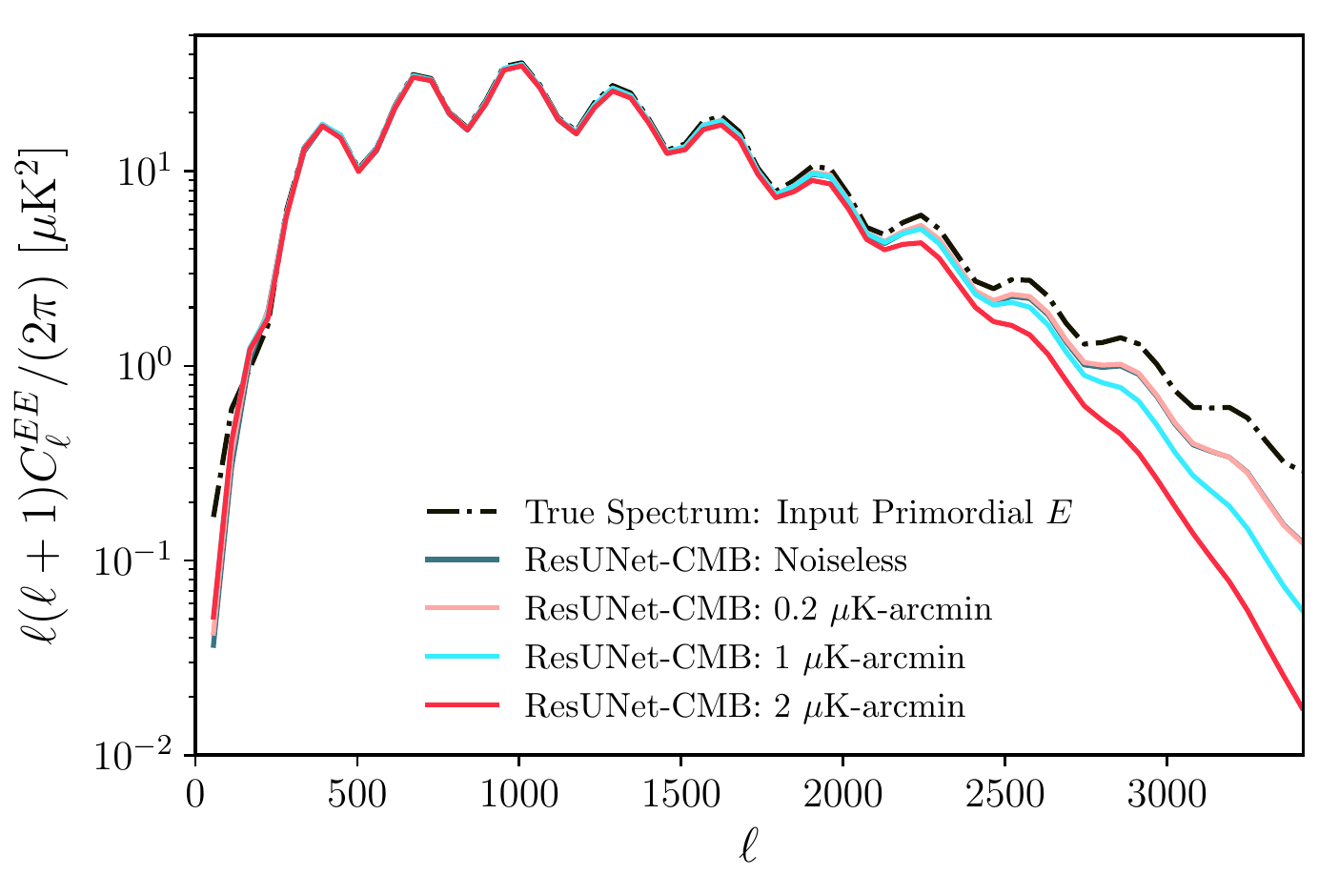}
    \caption{Power spectra of the primordial $E$-mode polarization maps $\biasmap{E}$ predicted by ResUNet-CMB  for each of four noise levels averaged over the 7000 realizations included in the prediction set, compared to the averaged power spectrum of the true primordial $E$ modes (black dashed-dot).
    }
    \label{fig:unle_ps}
\end{figure}

\subsection{Power Spectra}
Next, we compare the recovered pseudo-power spectra from the ResUNet-CMB prediction at each noise level to the truth power spectrum for each map. In Fig.~\ref{fig:tau_ps} we show a naively corrected $\langle C_\ell^{\biasmap{\tau} \biasmap{\tau}} \rangle$ pseudo-power spectrum for each noise level. To get the spectra we calculate the auto-power spectrum of each $\biasmap{\tau}$ map from the prediction set and average the results to find $\langle C^{\biasmap{\tau} \biasmap{\tau}}_{\ell} \rangle$. We then make a simple correction for power loss due to the window applied to each map by dividing the spectra by the mean of the squared cosine taper. The true input spectrum is the window-corrected average pseudo-power spectra of the noiseless truth maps $\tau(\nhat)$ in the prediction set.

As expected, reconstruction of $\tau$ by the ResUNet-CMB network is best at low $\ell$ (large angular scales) and degrades on small scales. As we increase instrumental noise, the fidelity of the reconstruction worsens on all scales. There is a larger decrease in the reconstructed power when going from the noiseless case to 1~$\mu$K-arcmin than there is between 1~$\mu$K-arcmin and 2~$\mu$K-arcmin.

Figure~\ref{fig:kappa_ps} and Fig.~\ref{fig:unle_ps} show the power spectra for the ResUNet-CMB reconstructions, $\biasmap{\kappa}$ and $\biasmap{E}$, respectively. For the $\biasmap{\kappa}$ spectra, the reconstruction quickly worsens with increasing instrument noise, especially on small angular scales.
Of the three ResUNet-CMB outputs, $\kappa$ sees the strongest scale dependence in its deviation from the input spectrum.

The reconstructed primordial $E$-mode map gives results closer to the input than both the $\tau$ and $\kappa$ ResUNet-CMB estimates across the entire range of scales we consider. This is not surprising, given that on the scales we consider the primordial $E$-mode polarization does not differ a great deal from the modulated and lensed $E$-mode polarization, the latter of which is just a linear combination of the input $Q$ and $U$ maps in the absence of noise. We find the noiseless $E$-mode power spectrum recovery to be greater than about 98\% for the $\ell$ range of $170<\ell<1400$.
At about $\ell=3000$ we find $E$-mode recovery of around 86\%. 
At higher noise levels, the predicted primordial $E$-mode spectrum falls significantly below the input spectrum on small angular scales. The impact of noise on the reconstruction of $E$ modes is smaller than that for $\kappa$ and $\tau$ on all scales.

We note that the results for our ResUNet-CMB lensing reconstruction match well with what is found in Ref.~\cite{Caldeira:2018ojb}. 
However, we find that our reconstructed $E$-mode power is closer to the input spectrum on all scales than what was found in Ref.~\cite{Caldeira:2018ojb}, though we have not been able to identify the cause of this difference. 

\begin{figure*}[t]
    \centering
    \includegraphics[width=\textwidth]{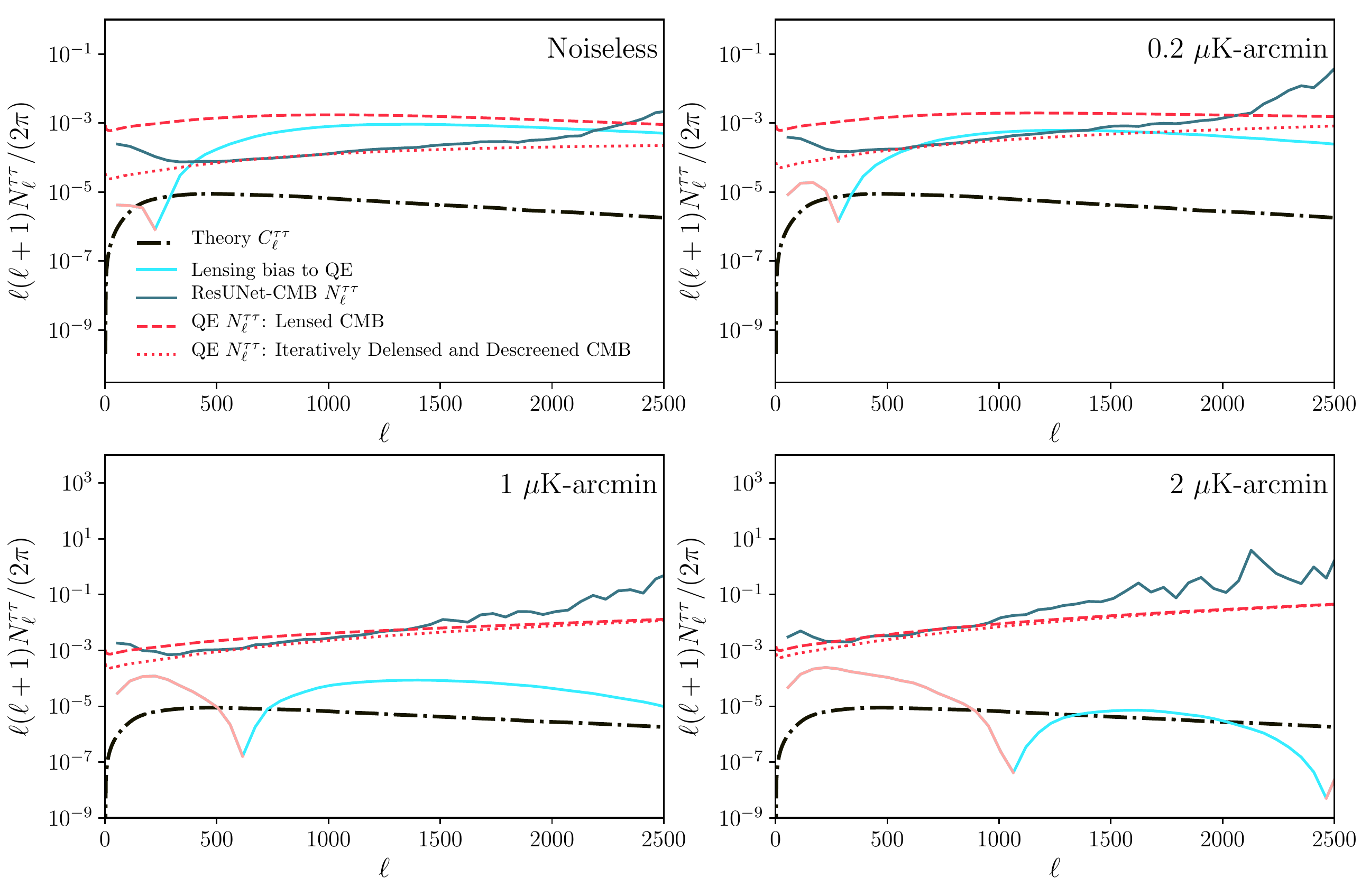}
    \caption{Reconstruction noise power spectra of ResUNet-CMB normalized prediction $\hat{\tau}$ as defined in Eq.~\eqref{eqn:noise_ps} (blue), standard quadratic estimator using lensed CMB spectra (QE; dashed red), and a quadratic estimator using  iteratively delensed and descreened CMB spectra (QE; dotted red). The expected lensing bias to the $EB$ quadratic estimator for $\tau$ (Eq.~\eqref{eqn:LensingBiasTh}) is plotted in cyan where it is positive and coral where negative.}
    \label{fig:nltt_wbias}
\end{figure*}

\subsection{Noise}
We can also compare the results of the ResUNet-CMB network to the standard quadratic estimator.  In order to do so, we need to define the reconstruction noise of the deep learning results.   The ResUNet-CMB prediction of the modulation field, after post-processing, is a biased result $\biasmap{\tau}$. Before calculating the reconstruction noise we first need to address this bias. Following treatment of the bias to the reconstructed lensing field from Ref.~\cite{Caldeira:2018ojb}, we define a quantity
\begin{equation}
    \label{eqn:map_normalization}
    A_{\ell} = \left[ \frac{\langle C_{\ell}^{\tau \utaumap}\rangle}{\langle C_{\ell}^{\tau \tau}\rangle} \right]^{-1} \, ,
\end{equation}
where $\langle \cdots  \rangle$ represents the average over the entire prediction set.  The quantity $A_\ell$ is used to rescale the biased output to recover an unbiased estimate of the optical depth fluctuations
\begin{equation}
    \label{eqn:norm_prediction}
    \hat{\tau}(\vect{\ell}) = A_{\ell} \utaumap(\vect{\ell}) \, .
\end{equation}
Estimates on real data using ResUNet-CMB would also require rescaling by the same quantity $A_\ell$.
The unbiased estimate $\hat{\tau}$ can be directly compared to the unbiased estimate from the quadratic estimator discussed in Section~\ref{sec:QE}.
The reconstruction noise spectrum can then be defined in a way analogous to the definition in Ref.~\cite{Caldeira:2018ojb},
\begin{equation}
    \label{eqn:noise_ps}
    N_{\ell}^{\tau \tau} =  \langle C_{\ell}^{\hat{\tau} \hat{\tau}} \rangle - \langle C_{\ell}^{\tau \tau} \rangle \, ,
\end{equation}
where we have used
\begin{equation}
    \label{eqn:unbiased_ps}
    \langle C_{\ell}^{\hat{\tau} \hat{\tau}} \rangle = A_{\ell}^{2} \langle C_{\ell}^{\biasmap{\tau} \biasmap{\tau}} \rangle \, .
\end{equation}
It is this definition of the reconstruction noise that we will compare to the quadratic estimator.

\begin{figure*}[t]
    \centering
    \includegraphics[width=\textwidth]{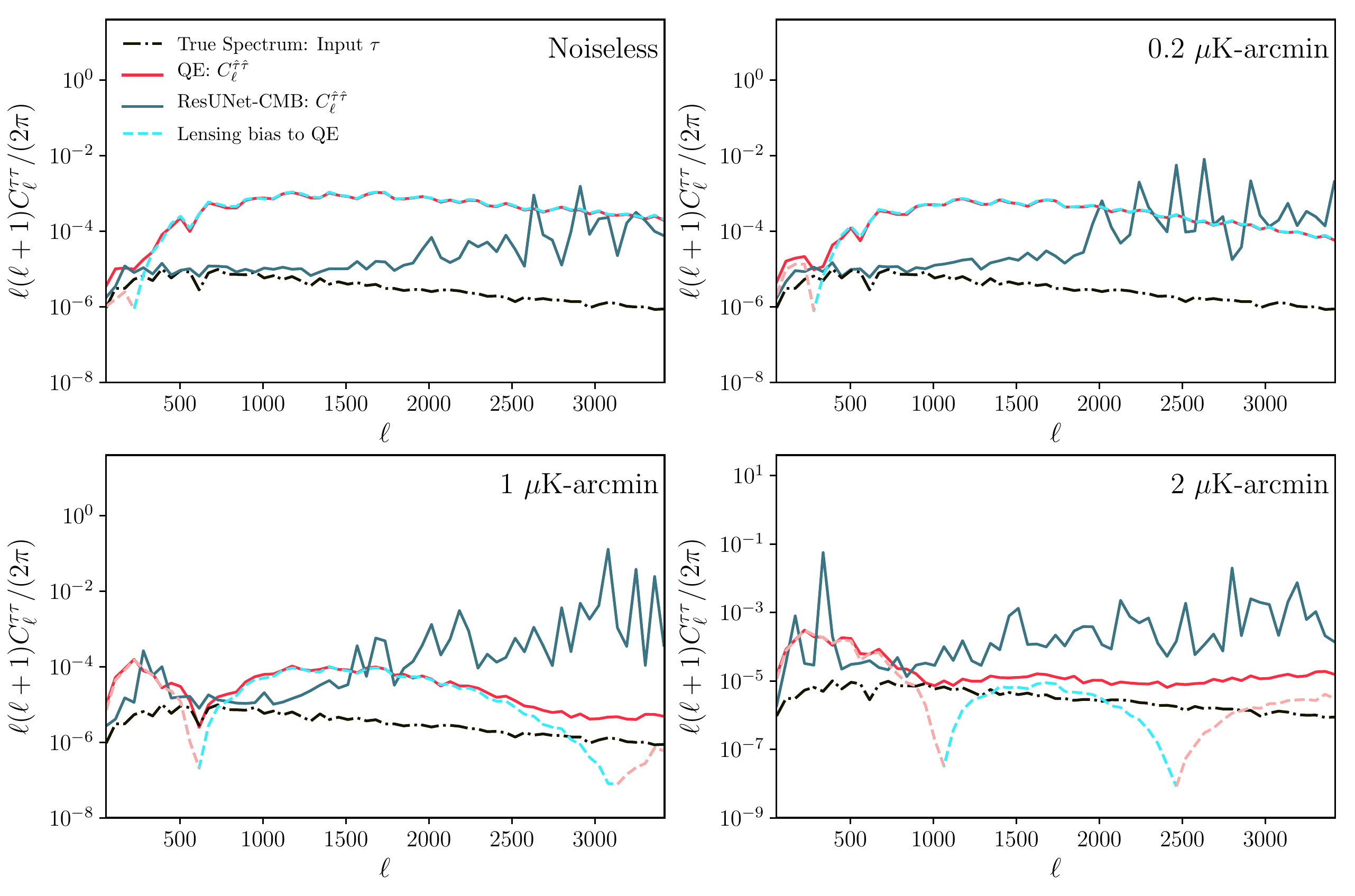}
    \caption{
    Power spectra of the patchy $\tau$ field reconstructed by ResUNet-CMB (blue) and the quadratic estimator (QE; red) computed from reconstructed maps that are averaged over a set of 7000 simulations with a fixed lensing convergence and patchy reionization field, compared with the
    true input $\tau$ signal power spectrum (black dash-dot) and the expected lensing bias to the quadratic estimator (cyan dashed where the bias is positive, coral dashed where the bias is negative). The sharp features in the power spectra
    are dependent on the random seeds chosen for the simulations. The quadratic estimator spectra
    were calculated from maps with no cosine taper applied, and we perform a mean-field subtraction on the estimate, $\hat{\tau}$.}
    \label{fig:bias_compare}
\end{figure*}

We calculate the reconstruction noise of the $EB$ quadratic estimator using Eq.~\eqref{eqn:Nltau_QE} (with $\ell_\mathrm{max}=6100$).
As discussed in Section~\ref{sec:QE}, the variance of the $EB$ estimator is increased in the presence of lensing, due to the $B$-mode polarization induced by lensing.  We therefore consider two versions of the quadratic estimator: one for which we used lensed CMB spectra, and another  for which we employ an iterative delensing and descreening procedure to the CMB spectra.

The iteratively delensed and descreened estimate of the patchy reionization reconstruction noise was calculated as follows. First, we calculate the effects of patchy reionization on the CMB polarization spectra.  We use these screened spectra to calculate an estimate of the lensing reconstruction noise using the iterative $EB$ estimator~\cite{Smith:2010gu}.  The $B$-mode polarization from the patchy reionization acts as an effective source of noise for lensing reconstruction, thereby increasing the lensing reconstruction noise as compared to a case without patchy reionization.  We use the resulting lensing reconstruction noise to compute delensed CMB spectra according to~Ref.~\cite{Green:2016cjr}.  The resulting delensed CMB spectra were used in the estimate of the patchy reionization reconstruction noise using the $EB$ estimator, with the residual lensing $B$-mode polarization acting as a source of noise for the patchy reionization reconstruction.  If we had stopped the procedure at this point, we would find that ResUNet-CMB outperformed this delensed quadratic estimator in the low noise cases we considered.  However, a lower patchy reionization reconstruction noise can be achieved by iteratively `descreening' the CMB to remove the best estimate of the effects of patchy reionization on the polarization, in a procedure analogous to delensing~\cite{Meerburg:2017lfh}.  We performed this iterative $EB$ descreening estimate, then used the residual effects of patchy reionization on the polarization spectra to compute an improved estimate of the lensing reconstruction noise using the iterative $EB$ estimator.  This resulted in a lower lensing reconstruction noise which was then used to calculate an improved estimate of delensed spectra.  We iterated this procedure of iterative $EB$ lensing reconstruction and delensing followed by iterative $EB$ patchy reionization reconstruction and descreening followed by iterative $EB$ lensing reconstruction and delensing, etc., to convergence.   The output after convergence of this procedure was used as our estimate of `Iteratively Delensed and Descreened' patchy reionization reconstruction noise.  Since the results of iterative delensing closely match the performance of the maximum likelihood lensing estimator~\cite{Hirata:2003ka,Smith:2010gu}, it should be expected that the iterative delensing and descreening procedure provides a good approximation of the noise resulting from the maximum likelihood estimate of simultaneous lensing and patchy reionization reconstruction, though the latter has not been demonstrated explicitly.

\begin{figure*}[t!]
    \centering
    \includegraphics[width=\textwidth]{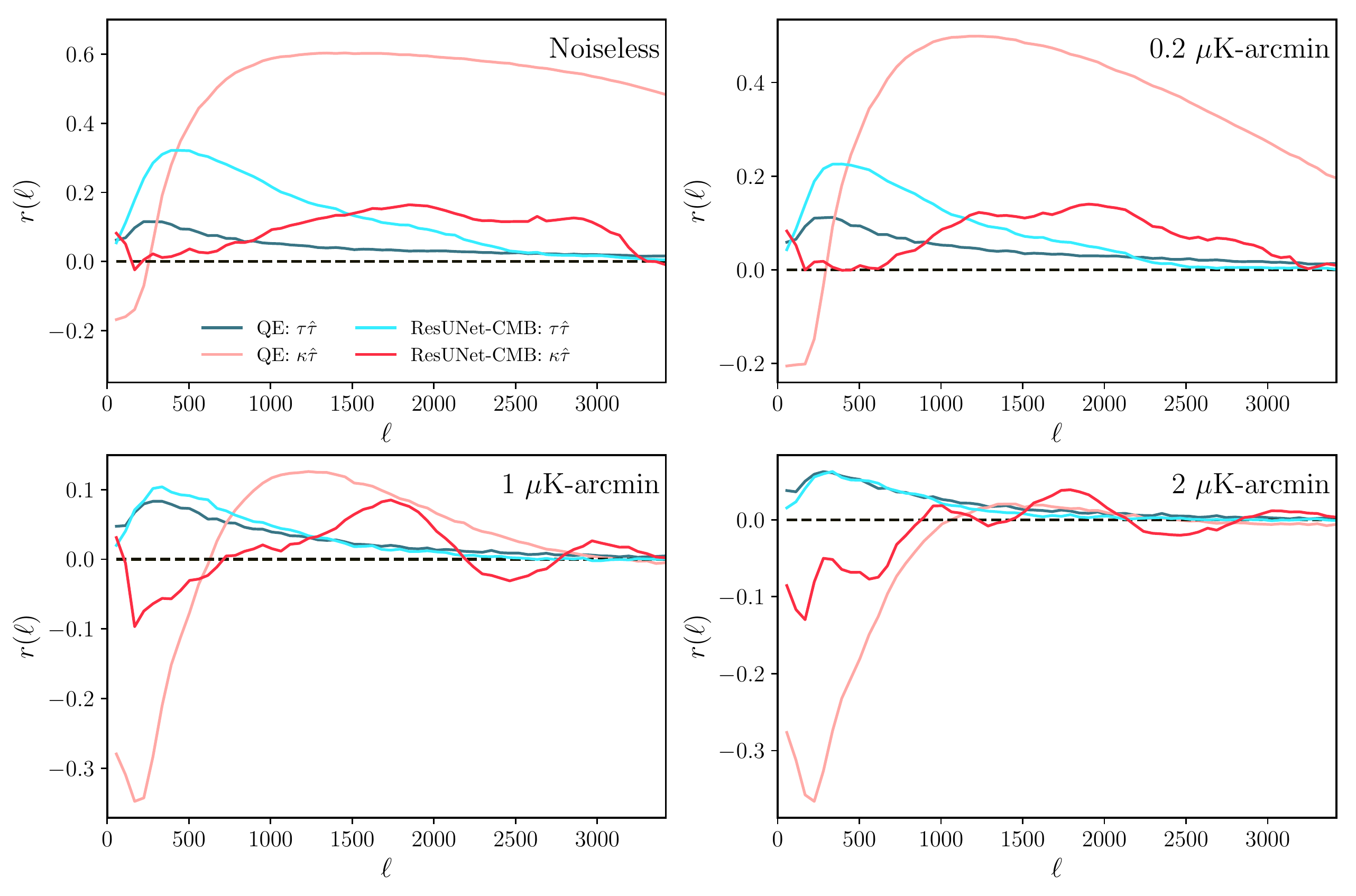}
    \caption{Correlation coefficients of estimates of the patchy reionization field $\hat{\tau}$ from the standard quadratic estimator (QE) with the true patchy reionization field $\tau$ (blue) and the lensing convergence field $\kappa$ (coral), and those from the ResUNet-CMB network with $\tau$ (cyan) and $\kappa$ (red), all averaged over 7000 simulations.
    The simulations for the quadratic estimator 
    did not use a cosine taper nor mean-field subtraction.
    }
    \label{fig:kt_corr_fac}
\end{figure*}

In addition to the extra variance due to lensing, the quadratic estimator also exhibits a bias to the reconstruction of the patchy reionization in the presence of lensing.  We calculate this bias according to Eq.~\eqref{eqn:LensingBiasTh}.
The bias-hardened estimator from Ref.~\cite{Su:2011ff} has only a percent-level increase in variance compared to the standard estimator, and so we present the reconstruction noise computed from the standard estimator. 

Figure~\ref{fig:nltt_wbias} shows the patchy reionization reconstruction noise power spectra for the quadratic estimator and the normalized ResUNet-CMB predictions for each noise level.  We also include in each panel the theoretical patchy reionization power spectrum and the bias on the standard quadratic estimator due to lensing.
For all four noise levels shown, the ResUNet-CMB predictions have a lower reconstruction noise than the quadratic estimator using lensed spectra in the range $200<\ell<1000$. The range of angular scales where ResUNet-CMB outperforms the standard quadratic estimator grows as the noise level decreases.
The ResUNet-CMB $N_{\ell}^{\tau \tau }$ matches well with the reconstruction noise of the quadratic estimator using iteratively delensed and descreened CMB spectra for a large $\ell$ range for all four noise levels.
The top two panels demonstrate that the patchy reconstruction noise from ResUNet-CMB is also smaller than the bias due to lensing that is expected in the standard quadratic estimator over a wide range of scales, and this range is wider at lower noise.
This behavior indicates that the ResUNet-CMB network learns to reconstruct the patchy $\tau$ field in a way that mitigates both the extra variance and the bias due to lensing.
The ResUNet-CMB network is therefore able to successfully reconstruct both the lensing convergence and patchy reionization fields simultaneously in a way that is nearly optimal.

\subsection{Bias}
In Fig.~\ref{fig:bias_compare} we show the reconstructed patchy reionization power spectrum from ResUNet-CMB and from the standard quadratic estimator for each noise level.  We implement the $EB$ quadratic estimator for $\tau$ as in Eq.~\eqref{eqn:PatchyTauRecon}, by using  \texttt{Symlens}\footnote{\url{https://github.com/simonsobs/symlens}}.  The results are computed by averaging over a set of 7000 simulated CMB maps with fixed $\kappa$ and $\tau$ maps; the random seed for $\kappa$ was set to 133 and for $\tau$ to 180. As expected, the quadratic estimator $C_{\ell}^{\hat{\tau} \hat{\tau}}$ matches well with the lensing bias computed from Eq.~\eqref{eqn:LensingBiasTh}. The ResUNet-CMB estimator on the other hand follows more closely the input signal power spectrum, which is most easily seen for the low noise cases. At low noise the ResUNet-CMB predictions deviate from the input spectrum at small angular scales ($\ell\gtrsim1500$), and for higher noise deviations appear even at larger scales.

To confirm that the ResUNet-CMB architecture is truly sensitive to the input $\tau$ and is not producing a spurious signal due to lensing or an artifact of the procedure, we calculate the correlation coefficients of each reconstruction according to
\begin{equation}\label{eq:corr_coeff}
    r_{Z \hat{\tau}}(\ell) =
    \frac{\langle C_{\ell}^{Z \hat{\tau}} \rangle}{\sqrt{\langle C_{\ell}^{Z Z} \rangle \langle C_{\ell}^{\hat{\tau} \hat{\tau}} \rangle }} \, ,
\end{equation}
where $Z=\kappa, \tau$.
In Fig.~\ref{fig:kt_corr_fac} we compare the correlation coefficients of the ResUNet-CMB predictions to those of the standard quadratic estimator. Note that our definition of the correlation coefficient in Eq.~\eqref{eq:corr_coeff} includes in the denominator the reconstruction noise on the patchy reionization field, and therefore these coefficients are always less than unity.
It can clearly be seen that for all noise levels and on almost all scales, the standard quadratic estimator produces an estimate for the optical depth perturbation $\hat{\tau}$ which is more strongly correlated with the lensing convergence $\kappa$ than it is with the true patchy reionization signal $\tau$.
The ResUNet-CMB estimate $\hat{\tau}$ has a larger correlation with the true patchy $\tau$ signal and smaller correlation with $\kappa$ than does the quadratic estimator for nearly all scales at each noise level.

Furthermore, the ResUNet-CMB patchy reionization estimate is more strongly correlated with the true signal than with the lensing convergence on large scales, considerably so at low noise levels.
While the quadratic estimator is dominated by the $\kappa$ signal on almost all scales, 
the ResUNet-CMB estimator appears to exhibit a lensing bias only on small scales for the two lowest noise levels, and a reduced bias for the largest noise cases shown.

For $\ell>2000$ in the 1~$\mu$K-arcmin and 2~$\mu$K-arcmin cases, the near zero positive correlation between $\hat{\tau}$ and the true patchy $\tau$ signal indicate ResUNet-CMB is unable to provide an accurate patchy reionization reconstruction on small scales at these noise levels. This might also explain similarity of the reconstructed patchy reionization spectra on small scales for these noise levels seen in Fig.~\ref{fig:tau_ps}.
\subsection{Null Test}

\begin{figure}[t]
    \centering
    \includegraphics[width=\linewidth]{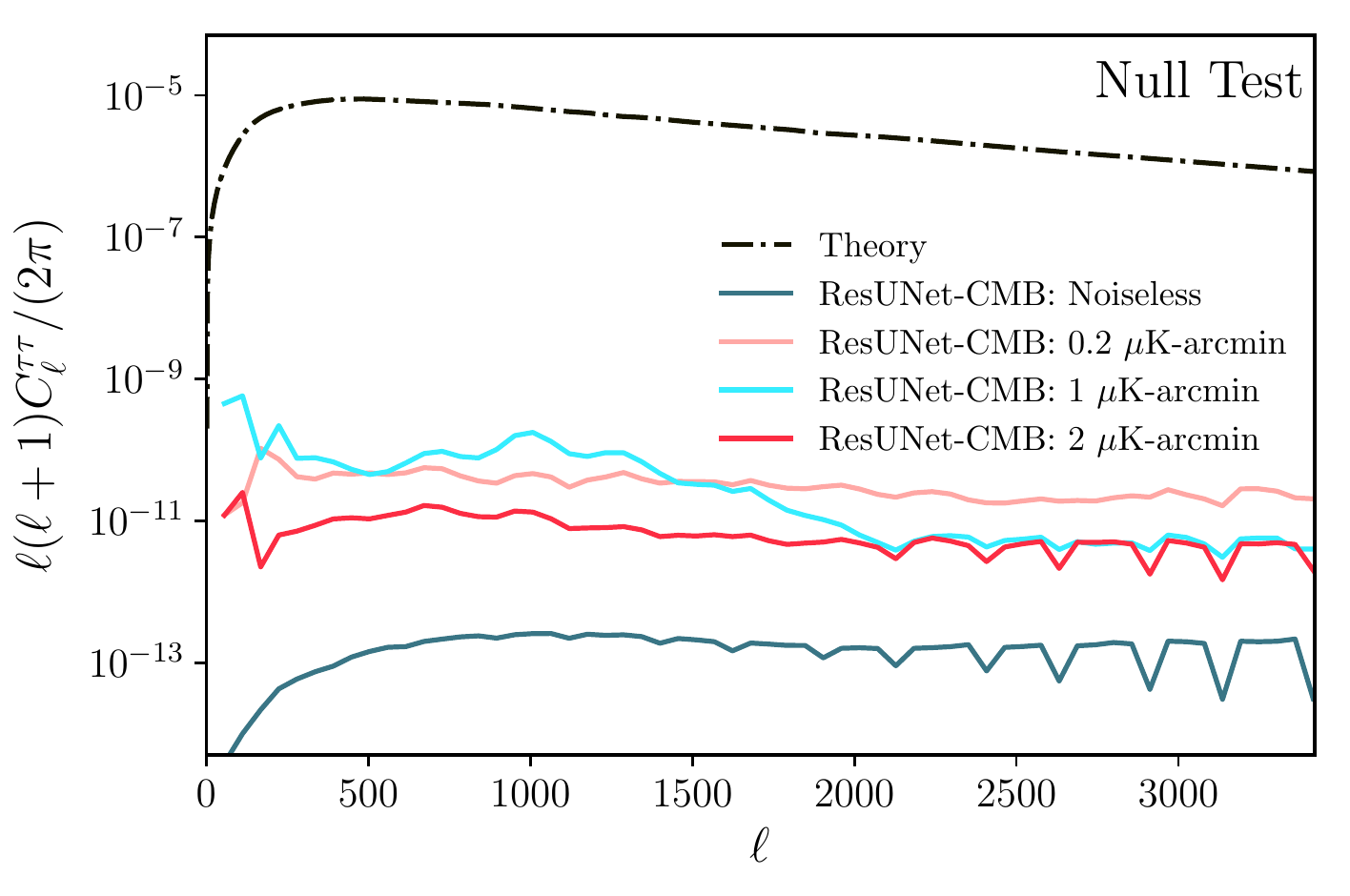}
    \caption{Null test showing the average power spectra of the patchy reionization field $\biasmap{\tau}$ predicted from fully trained ResUNet-CMB networks on a data set of 7000 simulations with $\kappa$ and $\tau$ set to zero.  The results of this figure should be compared to those in Fig.~\ref{fig:tau_ps}.  The theoretical patchy reionization spectrum that is used elsewhere in this work has been included in this plot (black dash-dot) to ease comparison, but for the null test shown in this figure, the true $C_\ell^{\tau\tau}$ vanishes.
    }
    \label{fig:null_ps}
\end{figure}

As a final check on the results of the ResUNet-CMB network, we perform a null test to ensure that no patchy reionization signal is reconstructed when none is present in the maps.
We implement this test by making predictions of $({\kappa},{\tau},\unlmap{E})$ with the fully trained networks when the unlensed and unmodulated polarization maps $(\unlmap{Q},\unlmap{U})$ (plus noise) are passed as the input. 
In Fig.~\ref{fig:null_ps}, we plot the power spectra $\langle C_{\ell}^{\biasmap{\tau} \biasmap{\tau}} \rangle$ for each noise level that results from this null test.  The spectra in Fig.~\ref{fig:null_ps} should be compared to the analogous spectra shown in Fig.~\ref{fig:tau_ps} where the patchy reionization signal was present.
We find that ResUNet-CMB reconstructs spectra $\langle C_{\ell}^{\biasmap{\tau} \biasmap{\tau}} \rangle$ that are much smaller for the null test than when a signal is present, especially on large angular scales where the reconstruction is most successful. In the noiseless case, the reconstructed spectrum $\langle C_{\ell}^{\biasmap{\tau} \biasmap{\tau}} \rangle$ in the null test is five to seven orders of magnitude smaller than the spectrum with patchy reionization included. 

We found that the ResUNet-CMB network was much less successful on this null test when trained without including unlensed and unmodulated maps in the training set.  In that case, the model tended to predict the presence of a patchy reionization signal with power spectra not much smaller than those shown in Fig.~\ref{fig:tau_ps}, even when no modulation was present in the maps.
The inclusion of some unlensed and unmodulated polarization maps in the data set during training prevents a spurious signal from being reconstructed when lensing and patchy reionization are absent.


\section{Conclusion} 
In this paper we described the construction of a deep learning network designed to reconstruct effects which alter the statistics of CMB fluctuations, the ResUNet-CMB.  We showed that this network is capable of the simultaneous reconstruction of a lensing map, a patchy reionization map, and a delensed and descreened polarization map.
The reconstruction of patchy reionization provided by ResUNet-CMB drastically reduced the lensing-induced bias present in the standard quadratic estimator, and it nearly matched the variance of an optimal iteratively delensed and descreened reconstruction scheme over a wide range of angular scales.

While the results from ResUNet-CMB are quite promising, we confirm the conclusion of other studies that with CMB polarization data alone, we are likely to be able to achieve only a statistical detection of the effects of patchy reionization~\cite{Su:2011ff}.
The success of ResUNet-CMB on the simultaneous reconstruction of two sources of secondary CMB fluctuations serves as a demonstration that machine learning tools can be usefully applied to the data expected from the next generation of CMB surveys.

The ResUNet-CMB network shows promising results, though improvements can still be made. The results of the network slightly under-performed idealized predictions for $({\kappa},{\tau},{E})$ at very low noise levels, especially on small scales.
In Ref.~\cite{Caldeira:2018ojb} it is suggested that the reconstruction of high-$\ell$ modes is limited by the signal-to-noise ratio at those small scales. For the case of patchy reionization this issue is of particular importance since its effects are much smaller than lensing.
This problem can potentially be partially mitigated by using higher resolution maps to include smaller physical scales, which should improve the reconstruction on all scales. 
Accommodating maps of a different resolution requires extensive restructuring of the network to ensure that features on all scales are being properly analyzed by the network while ensuring VRAM constraints are not exceeded.
Another avenue to consider would be to first train the network on patchy reionization maps with a magnified signal, then slowly decrease the signal magnitude until the fiducial amplitude is reached. This method, used in Ref.~\cite{Ciuca:2019oka} for a different purpose, could help with the reconstruction on scales dominated by noise. 
On the other hand, the noise levels where significant gains seem possible are much lower than those of currently planned CMB surveys, meaning that optimization in that very low noise regime may not lead to much practical improvement with forthcoming data.

One extension to our work that could improve patchy reionization reconstruction is to consider a broader set of input maps.  One simple addition is to include CMB temperature maps in addition to polarization. More realistic simulations would also include correlations between the patchy reionization field and the matter density, leading to correlations with the lensing field, galaxy counts, or other tracers of structure.  If such correlations were included, one could also consider adding maps of the cosmic infrared background, surveys of galaxies, or 21-cm intensity maps, the addition of which has been shown to improve patchy reionization reconstruction by other means~\cite{Feng:2018lwj,Feng:2018eal,Hotinli:2020csk}.

Another extension of our work 
would be treat more realistic simulations, for example by adding astrophysical foregrounds or anisotropic noise. Applications of machine learning to simulated foregrounds on the full sky are already showing promising results \cite{Petroff:2020fbf}. Addressing the complications presented by more realistic simulations is a necessary step if machine learning is to be usefully applied to the data collected by upcoming CMB surveys. 



We found a few modifications of standard machine learning techniques to be beneficial for our purposes.
We included a batch normalization layer in all of our residual connections, contrary to the standard implementation of residual connections~\cite{he2015deep, he2016identity}. 
We found that the inclusion of the batch normalization layer lowered validation loss across all outputs, thereby improving all reconstructions. The success of this change in our network prompts further questions about whether similar changes would be beneficial to other machine learning applications.
When training the network, we included in the training set some maps which had both lensing and patchy reionization set to zero.  We found this to be necessary for the trained network to successfully pass the null test for patchy reionization reconstruction.  Without the null maps in the training set, we found that the network returned non-zero estimates for the patchy reionization signal, even when no patchy reionization was present in the maps of the test set. This procedure of including a subset of maps with varying conditions into the training set can be explored for further possible improvements to the network.
The potential benefits of this modified subset training method should not be understated. Not only were the impacts to the quality of the reconstructions minimal, the change to the training also fixed a problem in the predictions of the trained network without requiring any modification to the architecture.

We did not explore the generalization of our predictions to maps generated with different cosmological parameters.
Further tests are required to determine how well the network is able to reconstruct patchy reionization and lensing in CMB maps generated assuming different cosmologies.
The success of the modified subset training provides a positive indication that a similar technique could be applied to improve the generalization capability of the network without much negative impact to the results.

In order to be useful for constraining cosmological parameters, the predictions made by any estimator need to be incorporated into a framework that is capable of rigorously quantifying uncertainties.  Once trained, making predictions with ResUNet-CMB is very computationally efficient, and so the network provides good candidate to be included in a Monte Carlo exploration of posteriors.  Another possibility would be to incorporate ResUNet-CMB into a Bayesian framework like that described in Ref.~\cite{Millea:2020cpw}.

Our results act as a proof of concept that machine learning is capable of the simultaneous reconstruction of two fields that lead to the distortion of the statistics of CMB fluctuations. 
It is natural to ask whether machine learning can also be usefully applied to the reconstruction of other sources of secondary CMB fluctuations, perhaps by including additional maps as input.
The ResUNet-CMB network described here is a useful model for such exploration, since new architecture aimed at similar tasks can easily be constructed by simply modifying the inputs and outputs of the publicly available network.

The work presented here adds to the list of examples that show machine learning is likely to play a positive role in the future of cosmology, especially when applied to the wealth of new data expected from forthcoming cosmological surveys.

\section*{Acknowledgments}
The authors would like to thank Alex van Engelen and Marius Millea for helpful conversations.  This work is supported by the US Department of Energy under grant no.~DE-SC0010129.  Computations were carried out on ManeFrame II, a shared high-performance computing cluster at Southern Methodist University.

\bibliographystyle{utphys}
\bibliography{bibliography}

\providecommand{\href}[2]{#2}\begingroup\raggedright\begin{thebibliography}{100}

\bibitem{Ade:2018sbj}
{\bfseries Simons Observatory} Collaboration, P.~Ade {\em et~al.}, ``{The
  Simons Observatory: Science goals and forecasts},''
  \href{http://dx.doi.org/10.1088/1475-7516/2019/02/056}{{\em JCAP} {\bfseries
  02} (2019) 056}, \href{http://arxiv.org/abs/1808.07445}{{\ttfamily
  arXiv:1808.07445 [astro-ph.CO]}}.

\bibitem{Aravena:2019tye}
M.~Aravena {\em et~al.}, ``{The CCAT-Prime Submillimeter Observatory},''
  \href{http://arxiv.org/abs/1909.02587}{{\ttfamily arXiv:1909.02587
  [astro-ph.IM]}}.

\bibitem{Abazajian:2016yjj}
{\bfseries CMB-S4} Collaboration, K.~N. Abazajian {\em et~al.}, ``{CMB-S4
  Science Book, First Edition},''
  \href{http://arxiv.org/abs/1610.02743}{{\ttfamily arXiv:1610.02743
  [astro-ph.CO]}}.

\bibitem{Hanany:2019lle}
{\bfseries NASA PICO} Collaboration, S.~Hanany {\em et~al.}, ``{PICO: Probe of
  Inflation and Cosmic Origins},''
  \href{http://arxiv.org/abs/1902.10541}{{\ttfamily arXiv:1902.10541
  [astro-ph.IM]}}.

\bibitem{Sehgal:2019ewc}
N.~Sehgal {\em et~al.}, ``{CMB-HD: An Ultra-Deep, High-Resolution
  Millimeter-Wave Survey Over Half the Sky},''
  \href{http://arxiv.org/abs/1906.10134}{{\ttfamily arXiv:1906.10134
  [astro-ph.CO]}}.

\bibitem{Aghanim:2007bt}
N.~Aghanim, S.~Majumdar, and J.~Silk, ``{Secondary anisotropies of the CMB},''
  \href{http://dx.doi.org/10.1088/0034-4885/71/6/066902}{{\em Rept. Prog.
  Phys.} {\bfseries 71} (2008) 066902},
  \href{http://arxiv.org/abs/0711.0518}{{\ttfamily arXiv:0711.0518
  [astro-ph]}}.

\bibitem{Lewis:2006fu}
A.~Lewis and A.~Challinor, ``{Weak gravitational lensing of the CMB},''
  \href{http://dx.doi.org/10.1016/j.physrep.2006.03.002}{{\em Phys. Rept.}
  {\bfseries 429} (2006) 1--65},
  \href{http://arxiv.org/abs/astro-ph/0601594}{{\ttfamily
  arXiv:astro-ph/0601594}}.

\bibitem{Zeldovich:1969ff}
{\relax Ya}.~B. Zeldovich and R.~A. Sunyaev, ``{The Interaction of Matter and
  Radiation in a Hot-Model Universe},''
\href{http://dx.doi.org/10.1007/BF00661821}{{\em Astrophys. Space Sci.}
  {\bfseries 4} (1969) 301--316}.

\bibitem{Sunyaev:1970er}
R.~A. Sunyaev and {\relax Ya}.~B. Zeldovich, ``{The Interaction of matter and
  radiation in the hot model of the universe},''
{\em Astrophys. Space Sci.} {\bfseries 7} (1970) 20--30.

\bibitem{Sunyaev:1972eq}
R.~A. Sunyaev and {\relax Ya}.~B. Zeldovich, ``{The Observations of relic
  radiation as a test of the nature of X-Ray radiation from the clusters of
  galaxies},''
{\em Comments Astrophys. Space Phys.} {\bfseries 4} (1972) 173--178.

\bibitem{Sunyaev:1980vz}
R.~A. Sunyaev and {\relax Ya}.~B. Zeldovich, ``{Microwave background radiation
  as a probe of the contemporary structure and history of the universe},''
\href{http://dx.doi.org/10.1146/annurev.aa.18.090180.002541}{{\em Ann. Rev.
  Astron. Astrophys.} {\bfseries 18} (1980) 537--560}.

\bibitem{Sazonov:1999zp}
S.~Y. Sazonov and R.~A. Sunyaev, ``{Microwave polarization in the direction of
  galaxy clusters induced by the CMB quadrupole anisotropy},''
  \href{http://dx.doi.org/10.1046/j.1365-8711.1999.02981.x}{{\em Mon. Not. Roy.
  Astron. Soc.} {\bfseries 310} (1999) 765--772},
\href{http://arxiv.org/abs/astro-ph/9903287}{{\ttfamily arXiv:astro-ph/9903287
  [astro-ph]}}.

\bibitem{1967ApJ...147...73S}
R.~K. {Sachs} and A.~M. {Wolfe}, ``{Perturbations of a Cosmological Model and
  Angular Variations of the Microwave Background},''
  \href{http://dx.doi.org/10.1086/148982}{{\em \apj} {\bfseries 147} (Jan.,
  1967) 73}.

\bibitem{1968Natur.217..511R}
M.~J. {Rees} and D.~W. {Sciama}, ``{Large-scale Density Inhomogeneities in the
  Universe},'' \href{http://dx.doi.org/10.1038/217511a0}{{\em \nat} {\bfseries
  217} (Feb., 1968) 511--516}.

\bibitem{1983Natur.302..315B}
M.~{Birkinshaw} and S.~F. {Gull}, ``{A test for transverse motions of clusters
  of galaxies},'' \href{http://dx.doi.org/10.1038/302315a0}{{\em \nat}
  {\bfseries 302} (Mar., 1983) 315--317}.

\bibitem{Smith:2016lnt}
K.~M. Smith and S.~Ferraro, ``{Detecting Patchy Reionization in the Cosmic
  Microwave Background},''
  \href{http://dx.doi.org/10.1103/PhysRevLett.119.021301}{{\em Phys. Rev.
  Lett.} {\bfseries 119} no.~2, (2017) 021301},
  \href{http://arxiv.org/abs/1607.01769}{{\ttfamily arXiv:1607.01769
  [astro-ph.CO]}}.

\bibitem{Louis:2017hoh}
T.~Louis, E.~F. Bunn, B.~Wandelt, and J.~Silk, ``{Measuring Polarized Emission
  in Clusters in the CMB S4 Era},''
  \href{http://dx.doi.org/10.1103/PhysRevD.96.123509}{{\em Phys. Rev. D}
  {\bfseries 96} no.~12, (2017) 123509},
  \href{http://arxiv.org/abs/1707.04102}{{\ttfamily arXiv:1707.04102
  [astro-ph.CO]}}.

\bibitem{Deutsch:2017cja}
A.-S. Deutsch, M.~C. Johnson, M.~M\"unchmeyer, and A.~Terrana, ``{Polarized
  Sunyaev Zel'dovich tomography},''
  \href{http://dx.doi.org/10.1088/1475-7516/2018/04/034}{{\em JCAP} {\bfseries
  04} (2018) 034}, \href{http://arxiv.org/abs/1705.08907}{{\ttfamily
  arXiv:1705.08907 [astro-ph.CO]}}.

\bibitem{Deutsch:2017ybc}
A.-S. Deutsch, E.~Dimastrogiovanni, M.~C. Johnson, M.~M\"unchmeyer, and
  A.~Terrana, ``{Reconstruction of the remote dipole and quadrupole fields from
  the kinetic Sunyaev Zel\textquoteright{}dovich and polarized Sunyaev
  Zel\textquoteright{}dovich effects},''
  \href{http://dx.doi.org/10.1103/PhysRevD.98.123501}{{\em Phys. Rev. D}
  {\bfseries 98} no.~12, (2018) 123501},
  \href{http://arxiv.org/abs/1707.08129}{{\ttfamily arXiv:1707.08129
  [astro-ph.CO]}}.

\bibitem{Meyers:2017rtf}
J.~Meyers, P.~D. Meerburg, A.~van Engelen, and N.~Battaglia, ``{Beyond CMB
  cosmic variance limits on reionization with the polarized
  Sunyaev-Zel\textquoteright{}dovich effect},''
  \href{http://dx.doi.org/10.1103/PhysRevD.97.103505}{{\em Phys. Rev. D}
  {\bfseries 97} no.~10, (2018) 103505},
  \href{http://arxiv.org/abs/1710.01708}{{\ttfamily arXiv:1710.01708
  [astro-ph.CO]}}.

\bibitem{Smith:2018bpn}
K.~M. Smith, M.~S. Madhavacheril, M.~M\"unchmeyer, S.~Ferraro, U.~Giri, and
  M.~C. Johnson, ``{KSZ tomography and the bispectrum},''
  \href{http://arxiv.org/abs/1810.13423}{{\ttfamily arXiv:1810.13423
  [astro-ph.CO]}}.

\bibitem{Hotinli:2018yyc}
S.~C. Hotinli, J.~Meyers, N.~Dalal, A.~H. Jaffe, M.~C. Johnson, J.~B. Mertens,
  M.~M\"unchmeyer, K.~M. Smith, and A.~van Engelen, ``{Transverse Velocities
  with the Moving Lens Effect},''
  \href{http://dx.doi.org/10.1103/PhysRevLett.123.061301}{{\em Phys. Rev.
  Lett.} {\bfseries 123} no.~6, (2019) 061301},
  \href{http://arxiv.org/abs/1812.03167}{{\ttfamily arXiv:1812.03167
  [astro-ph.CO]}}.

\bibitem{Yasini:2018rrl}
S.~Yasini, N.~Mirzatuny, and E.~Pierpaoli, ``{Pairwise Transverse Velocity
  Measurement with the Rees\textendash{}Sciama Effect},''
  \href{http://dx.doi.org/10.3847/2041-8213/ab0bfe}{{\em Astrophys. J. Lett.}
  {\bfseries 873} no.~2, (2019) L23},
  \href{http://arxiv.org/abs/1812.04241}{{\ttfamily arXiv:1812.04241
  [astro-ph.CO]}}.

\bibitem{Hotinli:2020ntd}
S.~C. Hotinli, M.~C. Johnson, and J.~Meyers, ``{Optimal filters for the moving
  lens effect},'' \href{http://arxiv.org/abs/2006.03060}{{\ttfamily
  arXiv:2006.03060 [astro-ph.CO]}}.

\bibitem{Hu:2001kj}
W.~Hu and T.~Okamoto, ``{Mass reconstruction with cmb polarization},''
  \href{http://dx.doi.org/10.1086/341110}{{\em Astrophys. J.} {\bfseries 574}
  (2002) 566--574}, \href{http://arxiv.org/abs/astro-ph/0111606}{{\ttfamily
  arXiv:astro-ph/0111606}}.

\bibitem{Okamoto:2003zw}
T.~Okamoto and W.~Hu, ``{CMB lensing reconstruction on the full sky},''
  \href{http://dx.doi.org/10.1103/PhysRevD.67.083002}{{\em Phys. Rev.}
  {\bfseries D67} (2003) 083002},
\href{http://arxiv.org/abs/astro-ph/0301031}{{\ttfamily arXiv:astro-ph/0301031
  [astro-ph]}}.

\bibitem{Aghanim:2018oex}
{\bfseries Planck} Collaboration, N.~Aghanim {\em et~al.}, ``{Planck 2018
  results. VIII. Gravitational lensing},''
  \href{http://dx.doi.org/10.1051/0004-6361/201833886}{{\em Astron. Astrophys.}
  {\bfseries 641} (2020) A8}, \href{http://arxiv.org/abs/1807.06210}{{\ttfamily
  arXiv:1807.06210 [astro-ph.CO]}}.

\bibitem{Wu:2019hek}
W.~Wu {\em et~al.}, ``{A Measurement of the Cosmic Microwave Background Lensing
  Potential and Power Spectrum from 500 deg$^2$ of SPTpol Temperature and
  Polarization Data},'' \href{http://dx.doi.org/10.3847/1538-4357/ab4186}{{\em
  Astrophys. J.} {\bfseries 884} (2019) 70},
  \href{http://arxiv.org/abs/1905.05777}{{\ttfamily arXiv:1905.05777
  [astro-ph.CO]}}.

\bibitem{Darwish:2020fwf}
O.~Darwish {\em et~al.}, ``{The Atacama Cosmology Telescope: A CMB lensing mass
  map over 2100 square degrees of sky and its cross-correlation with BOSS-CMASS
  galaxies},'' \href{http://arxiv.org/abs/2004.01139}{{\ttfamily
  arXiv:2004.01139 [astro-ph.CO]}}.

\bibitem{Hirata:2002jy}
C.~M. Hirata and U.~Seljak, ``{Analyzing weak lensing of the cosmic microwave
  background using the likelihood function},''
  \href{http://dx.doi.org/10.1103/PhysRevD.67.043001}{{\em Phys. Rev.}
  {\bfseries D67} (2003) 043001},
\href{http://arxiv.org/abs/astro-ph/0209489}{{\ttfamily arXiv:astro-ph/0209489
  [astro-ph]}}.

\bibitem{Hirata:2003ka}
C.~M. Hirata and U.~Seljak, ``{Reconstruction of lensing from the cosmic
  microwave background polarization},''
  \href{http://dx.doi.org/10.1103/PhysRevD.68.083002}{{\em Phys. Rev. D}
  {\bfseries 68} (2003) 083002},
  \href{http://arxiv.org/abs/astro-ph/0306354}{{\ttfamily
  arXiv:astro-ph/0306354}}.

\bibitem{Smith:2010gu}
K.~M. Smith, D.~Hanson, M.~LoVerde, C.~M. Hirata, and O.~Zahn, ``{Delensing CMB
  Polarization with External Datasets},''
  \href{http://dx.doi.org/10.1088/1475-7516/2012/06/014}{{\em JCAP} {\bfseries
  06} (2012) 014}, \href{http://arxiv.org/abs/1010.0048}{{\ttfamily
  arXiv:1010.0048 [astro-ph.CO]}}.

\bibitem{Namikawa:2012pe}
T.~Namikawa, D.~Hanson, and R.~Takahashi, ``{Bias-Hardened CMB Lensing},''
  \href{http://dx.doi.org/10.1093/mnras/stt195}{{\em Mon. Not. Roy. Astron.
  Soc.} {\bfseries 431} (2013) 609--620},
  \href{http://arxiv.org/abs/1209.0091}{{\ttfamily arXiv:1209.0091
  [astro-ph.CO]}}.

\bibitem{Kamionkowski:1996ks}
M.~Kamionkowski, A.~Kosowsky, and A.~Stebbins, ``{Statistics of cosmic
  microwave background polarization},''
  \href{http://dx.doi.org/10.1103/PhysRevD.55.7368}{{\em Phys. Rev. D}
  {\bfseries 55} (1997) 7368--7388},
  \href{http://arxiv.org/abs/astro-ph/9611125}{{\ttfamily
  arXiv:astro-ph/9611125}}.

\bibitem{Seljak:1996gy}
U.~Seljak and M.~Zaldarriaga, ``{Signature of gravity waves in polarization of
  the microwave background},''
  \href{http://dx.doi.org/10.1103/PhysRevLett.78.2054}{{\em Phys. Rev. Lett.}
  {\bfseries 78} (1997) 2054--2057},
  \href{http://arxiv.org/abs/astro-ph/9609169}{{\ttfamily
  arXiv:astro-ph/9609169}}.

\bibitem{Zaldarriaga:1998ar}
M.~Zaldarriaga and U.~Seljak, ``{Gravitational lensing effect on cosmic
  microwave background polarization},''
  \href{http://dx.doi.org/10.1103/PhysRevD.58.023003}{{\em Phys. Rev. D}
  {\bfseries 58} (1998) 023003},
  \href{http://arxiv.org/abs/astro-ph/9803150}{{\ttfamily
  arXiv:astro-ph/9803150}}.

\bibitem{Lewis:2001hp}
A.~Lewis, A.~Challinor, and N.~Turok, ``{Analysis of CMB polarization on an
  incomplete sky},'' \href{http://dx.doi.org/10.1103/PhysRevD.65.023505}{{\em
  Phys. Rev. D} {\bfseries 65} (2002) 023505},
  \href{http://arxiv.org/abs/astro-ph/0106536}{{\ttfamily
  arXiv:astro-ph/0106536}}.

\bibitem{Knox:2002pe}
L.~Knox and Y.-S. Song, ``{A Limit on the detectability of the energy scale of
  inflation},'' \href{http://dx.doi.org/10.1103/PhysRevLett.89.011303}{{\em
  Phys. Rev. Lett.} {\bfseries 89} (2002) 011303},
  \href{http://arxiv.org/abs/astro-ph/0202286}{{\ttfamily
  arXiv:astro-ph/0202286}}.

\bibitem{Kesden:2002ku}
M.~Kesden, A.~Cooray, and M.~Kamionkowski, ``{Separation of gravitational wave
  and cosmic shear contributions to cosmic microwave background
  polarization},'' \href{http://dx.doi.org/10.1103/PhysRevLett.89.011304}{{\em
  Phys. Rev. Lett.} {\bfseries 89} (2002) 011304},
  \href{http://arxiv.org/abs/astro-ph/0202434}{{\ttfamily
  arXiv:astro-ph/0202434}}.

\bibitem{Seljak:2003pn}
U.~Seljak and C.~M. Hirata, ``{Gravitational lensing as a contaminant of the
  gravity wave signal in CMB},''
  \href{http://dx.doi.org/10.1103/PhysRevD.69.043005}{{\em Phys. Rev. D}
  {\bfseries 69} (2004) 043005},
  \href{http://arxiv.org/abs/astro-ph/0310163}{{\ttfamily
  arXiv:astro-ph/0310163}}.

\bibitem{Green:2016cjr}
D.~Green, J.~Meyers, and A.~van Engelen, ``{CMB Delensing Beyond the B
  Modes},'' \href{http://dx.doi.org/10.1088/1475-7516/2017/12/005}{{\em JCAP}
  {\bfseries 1712} no.~12, (2017) 005},
\href{http://arxiv.org/abs/1609.08143}{{\ttfamily arXiv:1609.08143
  [astro-ph.CO]}}.

\bibitem{Baumann:2015rya}
D.~Baumann, D.~Green, J.~Meyers, and B.~Wallisch, ``{Phases of New Physics in
  the CMB},'' \href{http://dx.doi.org/10.1088/1475-7516/2016/01/007}{{\em JCAP}
  {\bfseries 01} (2016) 007}, \href{http://arxiv.org/abs/1508.06342}{{\ttfamily
  arXiv:1508.06342 [astro-ph.CO]}}.

\bibitem{Coulton:2019odk}
W.~R. Coulton, P.~D. Meerburg, D.~G. Baker, S.~Hotinli, A.~J. Duivenvoorden,
  and A.~van Engelen, ``{Minimizing gravitational lensing contributions to the
  primordial bispectrum covariance},''
  \href{http://dx.doi.org/10.1103/PhysRevD.101.123504}{{\em Phys. Rev. D}
  {\bfseries 101} no.~12, (2020) 123504},
  \href{http://arxiv.org/abs/1912.07619}{{\ttfamily arXiv:1912.07619
  [astro-ph.CO]}}.

\bibitem{Abazajian:2019eic}
K.~Abazajian {\em et~al.}, ``{CMB-S4 Science Case, Reference Design, and
  Project Plan},'' \href{http://arxiv.org/abs/1907.04473}{{\ttfamily
  arXiv:1907.04473 [astro-ph.IM]}}.

\bibitem{Abazajian:2020dmr}
{\bfseries CMB-S4} Collaboration, K.~Abazajian {\em et~al.}, ``{CMB-S4:
  Forecasting Constraints on Primordial Gravitational Waves},''
  \href{http://arxiv.org/abs/2008.12619}{{\ttfamily arXiv:2008.12619
  [astro-ph.CO]}}.

\bibitem{Millea:2017fyd}
M.~Millea, E.~Anderes, and B.~D. Wandelt, ``{Bayesian delensing of CMB
  temperature and polarization},''
  \href{http://dx.doi.org/10.1103/PhysRevD.100.023509}{{\em Phys. Rev. D}
  {\bfseries 100} no.~2, (2019) 023509},
  \href{http://arxiv.org/abs/1708.06753}{{\ttfamily arXiv:1708.06753
  [astro-ph.CO]}}.

\bibitem{Millea:2020cpw}
M.~Millea, E.~Anderes, and B.~D. Wandelt, ``{Bayesian delensing delight:
  sampling-based inference of the primordial CMB and gravitational lensing},''
  \href{http://arxiv.org/abs/2002.00965}{{\ttfamily arXiv:2002.00965
  [astro-ph.CO]}}.

\bibitem{Diego-Palazuelos:2020lme}
P.~Diego-Palazuelos, P.~Vielva, E.~Mart\'\i{}nez-Gonz\'alez, and R.~Barreiro,
  ``{Comparison of delensing methodologies and assessment of the delensing
  capabilities of future experiments},''
  \href{http://arxiv.org/abs/2006.12935}{{\ttfamily arXiv:2006.12935
  [astro-ph.CO]}}.

\bibitem{Sherwin:2015baa}
B.~D. Sherwin and M.~Schmittfull, ``{Delensing the CMB with the Cosmic Infrared
  Background},'' \href{http://dx.doi.org/10.1103/PhysRevD.92.043005}{{\em Phys.
  Rev. D} {\bfseries 92} no.~4, (2015) 043005},
  \href{http://arxiv.org/abs/1502.05356}{{\ttfamily arXiv:1502.05356
  [astro-ph.CO]}}.

\bibitem{Sehgal:2016eag}
N.~Sehgal, M.~S. Madhavacheril, B.~Sherwin, and A.~van Engelen, ``{Internal
  Delensing of Cosmic Microwave Background Acoustic Peaks},''
  \href{http://dx.doi.org/10.1103/PhysRevD.95.103512}{{\em Phys. Rev. D}
  {\bfseries 95} no.~10, (2017) 103512},
  \href{http://arxiv.org/abs/1612.03898}{{\ttfamily arXiv:1612.03898
  [astro-ph.CO]}}.

\bibitem{Barkana:2000fd}
R.~Barkana and A.~Loeb, ``{In the beginning: The First sources of light and the
  reionization of the Universe},''
  \href{http://dx.doi.org/10.1016/S0370-1573(01)00019-9}{{\em Phys. Rept.}
  {\bfseries 349} (2001) 125--238},
  \href{http://arxiv.org/abs/astro-ph/0010468}{{\ttfamily
  arXiv:astro-ph/0010468}}.

\bibitem{Wyithe:2002qu}
J.~B. Wyithe and A.~Loeb, ``{Reionization of hydrogen and helium by early stars
  and quasars},'' \href{http://dx.doi.org/10.1086/367721}{{\em Astrophys. J.}
  {\bfseries 586} (2003) 693--708},
  \href{http://arxiv.org/abs/astro-ph/0209056}{{\ttfamily
  arXiv:astro-ph/0209056}}.

\bibitem{Barkana:2003qk}
R.~Barkana and A.~Loeb, ``{Unusually large fluctuations in the statistics of
  galaxy formation at high redshift},''
  \href{http://dx.doi.org/10.1086/421079}{{\em Astrophys. J.} {\bfseries 609}
  (2004) 474--481}, \href{http://arxiv.org/abs/astro-ph/0310338}{{\ttfamily
  arXiv:astro-ph/0310338}}.

\bibitem{Furlanetto:2004nh}
S.~Furlanetto, M.~Zaldarriaga, and L.~Hernquist, ``{The Growth of HII regions
  during reionization},'' \href{http://dx.doi.org/10.1086/423025}{{\em
  Astrophys. J.} {\bfseries 613} (2004) 1--15},
  \href{http://arxiv.org/abs/astro-ph/0403697}{{\ttfamily
  arXiv:astro-ph/0403697}}.

\bibitem{McQuinn:2006et}
M.~McQuinn, A.~Lidz, O.~Zahn, S.~Dutta, L.~Hernquist, and M.~Zaldarriaga,
  ``{The Morphology of HII Regions during Reionization},''
  \href{http://dx.doi.org/10.1111/j.1365-2966.2007.11489.x}{{\em Mon. Not. Roy.
  Astron. Soc.} {\bfseries 377} (2007) 1043--1063},
  \href{http://arxiv.org/abs/astro-ph/0610094}{{\ttfamily
  arXiv:astro-ph/0610094}}.

\bibitem{Mesinger:2007pd}
A.~Mesinger and S.~Furlanetto, ``{Efficient Simulations of Early Structure
  Formation and Reionization},'' \href{http://dx.doi.org/10.1086/521806}{{\em
  Astrophys. J.} {\bfseries 669} (2007) 663},
  \href{http://arxiv.org/abs/0704.0946}{{\ttfamily arXiv:0704.0946
  [astro-ph]}}.

\bibitem{Battaglia:2012id}
N.~Battaglia, H.~Trac, R.~Cen, and A.~Loeb, ``{Reionization on Large Scales I:
  A Parametric Model Constructed from Radiation-Hydrodynamic Simulations},''
  \href{http://dx.doi.org/10.1088/0004-637X/776/2/81}{{\em Astrophys. J.}
  {\bfseries 776} (2013) 81}, \href{http://arxiv.org/abs/1211.2821}{{\ttfamily
  arXiv:1211.2821 [astro-ph.CO]}}.

\bibitem{Mitra:2016olz}
S.~Mitra, T.~R. Choudhury, and A.~Ferrara, ``{Cosmic reionization after Planck
  II: contribution from quasars},''
  \href{http://dx.doi.org/10.1093/mnras/stx2443}{{\em Mon. Not. Roy. Astron.
  Soc.} {\bfseries 473} no.~1, (2018) 1416--1425},
  \href{http://arxiv.org/abs/1606.02719}{{\ttfamily arXiv:1606.02719
  [astro-ph.CO]}}.

\bibitem{Dayal:2018hft}
P.~Dayal and A.~Ferrara, ``{Early galaxy formation and its large-scale
  effects},'' \href{http://dx.doi.org/10.1016/j.physrep.2018.10.002}{{\em Phys.
  Rept.} {\bfseries 780-782} (2018) 1--64},
  \href{http://arxiv.org/abs/1809.09136}{{\ttfamily arXiv:1809.09136
  [astro-ph.GA]}}.

\bibitem{Hu:1999vq}
W.~Hu, ``{Reionization revisited: secondary cmb anisotropies and
  polarization},'' \href{http://dx.doi.org/10.1086/308279}{{\em Astrophys. J.}
  {\bfseries 529} (2000) 12},
  \href{http://arxiv.org/abs/astro-ph/9907103}{{\ttfamily
  arXiv:astro-ph/9907103}}.

\bibitem{Santos:2003jb}
M.~G. Santos, A.~Cooray, Z.~Haiman, L.~Knox, and C.-P. Ma, ``{Small - scale CMB
  temperature and polarization anisotropies due to patchy reionization},''
  \href{http://dx.doi.org/10.1086/378772}{{\em Astrophys. J.} {\bfseries 598}
  (2003) 756--766}, \href{http://arxiv.org/abs/astro-ph/0305471}{{\ttfamily
  arXiv:astro-ph/0305471}}.

\bibitem{Zahn:2005fn}
O.~Zahn, M.~Zaldarriaga, L.~Hernquist, and M.~McQuinn, ``{The Influence of
  non-uniform reionization on the CMB},''
  \href{http://dx.doi.org/10.1086/431947}{{\em Astrophys. J.} {\bfseries 630}
  (2005) 657--666}, \href{http://arxiv.org/abs/astro-ph/0503166}{{\ttfamily
  arXiv:astro-ph/0503166}}.

\bibitem{McQuinn:2005ce}
M.~McQuinn, S.~R. Furlanetto, L.~Hernquist, O.~Zahn, and M.~Zaldarriaga, ``{The
  Kinetic Sunyaev-Zel'dovich effect from reionization},''
  \href{http://dx.doi.org/10.1086/432049}{{\em Astrophys. J.} {\bfseries 630}
  (2005) 643--656}, \href{http://arxiv.org/abs/astro-ph/0504189}{{\ttfamily
  arXiv:astro-ph/0504189}}.

\bibitem{Dore:2007bz}
O.~Dore, G.~Holder, M.~Alvarez, I.~T. Iliev, G.~Mellema, U.-L. Pen, and P.~R.
  Shapiro, ``{The Signature of Patchy Reionization in the Polarization
  Anisotropy of the CMB},''
  \href{http://dx.doi.org/10.1103/PhysRevD.76.043002}{{\em Phys. Rev. D}
  {\bfseries 76} (2007) 043002},
  \href{http://arxiv.org/abs/astro-ph/0701784}{{\ttfamily
  arXiv:astro-ph/0701784}}.

\bibitem{Dvorkin:2008tf}
C.~Dvorkin and K.~M. Smith, ``{Reconstructing Patchy Reionization from the
  Cosmic Microwave Background},''
  \href{http://dx.doi.org/10.1103/PhysRevD.79.043003}{{\em Phys. Rev. D}
  {\bfseries 79} (2009) 043003},
  \href{http://arxiv.org/abs/0812.1566}{{\ttfamily arXiv:0812.1566
  [astro-ph]}}.

\bibitem{Dvorkin:2009ah}
C.~Dvorkin, W.~Hu, and K.~M. Smith, ``{B-mode CMB Polarization from Patchy
  Screening during Reionization},''
  \href{http://dx.doi.org/10.1103/PhysRevD.79.107302}{{\em Phys. Rev. D}
  {\bfseries 79} (2009) 107302},
  \href{http://arxiv.org/abs/0902.4413}{{\ttfamily arXiv:0902.4413
  [astro-ph.CO]}}.

\bibitem{Battaglia:2012im}
N.~Battaglia, A.~Natarajan, H.~Trac, R.~Cen, and A.~Loeb, ``{Reionization on
  Large Scales III: Predictions for Low-$\ell$ Cosmic Microwave Background
  Polarization and High-$\ell$ Kinetic Sunyaev-Zel'dovich Observables},''
  \href{http://dx.doi.org/10.1088/0004-637X/776/2/83}{{\em Astrophys. J.}
  {\bfseries 776} (2013) 83}, \href{http://arxiv.org/abs/1211.2832}{{\ttfamily
  arXiv:1211.2832 [astro-ph.CO]}}.

\bibitem{Park:2013mv}
H.~Park, P.~R. Shapiro, E.~Komatsu, I.~T. Iliev, K.~Ahn, and G.~Mellema, ``{The
  Kinetic Sunyaev-Zel'dovich effect as a probe of the physics of cosmic
  reionization: the effect of self-regulated reionization},''
  \href{http://dx.doi.org/10.1088/0004-637X/769/2/93}{{\em Astrophys. J.}
  {\bfseries 769} (2013) 93}, \href{http://arxiv.org/abs/1301.3607}{{\ttfamily
  arXiv:1301.3607 [astro-ph.CO]}}.

\bibitem{Alvarez:2015xzu}
M.~A. Alvarez, ``{The Kinetic Sunyaev\textendash{}Zel\textquoteright{}dovich
  Effect From Reionization: Simulated Full-sky Maps at Arcminute Resolution},''
  \href{http://dx.doi.org/10.3847/0004-637X/824/2/118}{{\em Astrophys. J.}
  {\bfseries 824} no.~2, (2016) 118},
  \href{http://arxiv.org/abs/1511.02846}{{\ttfamily arXiv:1511.02846
  [astro-ph.CO]}}.

\bibitem{Paul:2020fio}
S.~Paul, S.~Mukherjee, and T.~R. Choudhury, ``{Inevitable imprints of patchy
  reionization on the cosmic microwave background anisotropy},''
  \href{http://dx.doi.org/10.1093/mnras/staa3221}{{\em Mon. Not. Roy. Astron.
  Soc.} {\bfseries 500} no.~1, (2020) 232--246},
  \href{http://arxiv.org/abs/2005.05327}{{\ttfamily arXiv:2005.05327
  [astro-ph.CO]}}.

\bibitem{Choudhury:2020kzh}
T.~R. Choudhury, S.~Mukherjee, and S.~Paul, ``{CMB constraints on a physical
  model of reionization},'' \href{http://arxiv.org/abs/2007.03705}{{\ttfamily
  arXiv:2007.03705 [astro-ph.CO]}}.

\bibitem{Mukherjee:2019zlb}
S.~Mukherjee, S.~Paul, and T.~R. Choudhury, ``{Is patchy reionization an
  obstacle in detecting the primordial gravitational wave signal?},''
  \href{http://dx.doi.org/10.1093/mnras/stz1002}{{\em Mon. Not. Roy. Astron.
  Soc.} {\bfseries 486} no.~2, (2019) 2042--2049},
  \href{http://arxiv.org/abs/1903.01994}{{\ttfamily arXiv:1903.01994
  [astro-ph.CO]}}.

\bibitem{Su:2011ff}
M.~Su, A.~P. Yadav, M.~McQuinn, J.~Yoo, and M.~Zaldarriaga, ``{An Improved
  Forecast of Patchy Reionization Reconstruction with CMB},''
  \href{http://arxiv.org/abs/1106.4313}{{\ttfamily arXiv:1106.4313
  [astro-ph.CO]}}.

\bibitem{Gluscevic:2012qv}
V.~Gluscevic, M.~Kamionkowski, and D.~Hanson, ``{Patchy Screening of the Cosmic
  Microwave Background by Inhomogeneous Reionization},''
  \href{http://dx.doi.org/10.1103/PhysRevD.87.047303}{{\em Phys. Rev. D}
  {\bfseries 87} no.~4, (2013) 047303},
  \href{http://arxiv.org/abs/1210.5507}{{\ttfamily arXiv:1210.5507
  [astro-ph.CO]}}.

\bibitem{Namikawa:2017uke}
T.~Namikawa, ``{Constraints on Patchy Reionization from Planck CMB Temperature
  Trispectrum},'' \href{http://dx.doi.org/10.1103/PhysRevD.97.063505}{{\em
  Phys. Rev. D} {\bfseries 97} no.~6, (2018) 063505},
  \href{http://arxiv.org/abs/1711.00058}{{\ttfamily arXiv:1711.00058
  [astro-ph.CO]}}.

\bibitem{Roy:2018gcv}
A.~Roy, A.~Lapi, D.~Spergel, and C.~Baccigalupi, ``{Observing patchy
  reionization with future CMB polarization experiments},''
  \href{http://dx.doi.org/10.1088/1475-7516/2018/05/014}{{\em JCAP} {\bfseries
  05} (2018) 014}, \href{http://arxiv.org/abs/1801.02393}{{\ttfamily
  arXiv:1801.02393 [astro-ph.CO]}}.

\bibitem{Caldeira:2018ojb}
J.~Caldeira, W.~Wu, B.~Nord, C.~Avestruz, S.~Trivedi, and K.~Story, ``{DeepCMB:
  Lensing Reconstruction of the Cosmic Microwave Background with Deep Neural
  Networks},'' \href{http://dx.doi.org/10.1016/j.ascom.2019.100307}{{\em
  Astron. Comput.} {\bfseries 28} (2019) 100307},
  \href{http://arxiv.org/abs/1810.01483}{{\ttfamily arXiv:1810.01483
  [astro-ph.CO]}}.

\bibitem{Petroff:2020fbf}
M.~A. Petroff, G.~E. Addison, C.~L. Bennett, and J.~L. Weiland, ``{Full-sky
  Cosmic Microwave Background Foreground Cleaning Using Machine Learning},''
  \href{http://arxiv.org/abs/2004.11507}{{\ttfamily arXiv:2004.11507
  [astro-ph.CO]}}.

\bibitem{Munchmeyer:2019kng}
M.~M\"unchmeyer and K.~M. Smith, ``{Fast Wiener filtering of CMB maps with
  Neural Networks},'' \href{http://arxiv.org/abs/1905.05846}{{\ttfamily
  arXiv:1905.05846 [astro-ph.CO]}}.

\bibitem{Gupta:2020him}
N.~Gupta and C.~Reichardt, ``{Mass Estimation of Galaxy Clusters with Deep
  Learning II: CMB Cluster Lensing},''
  \href{http://arxiv.org/abs/2005.13985}{{\ttfamily arXiv:2005.13985
  [astro-ph.CO]}}.

\bibitem{Gupta:2020yvd}
N.~Gupta and C.~L. Reichardt, ``{Mass Estimation of Galaxy Clusters with Deep
  Learning I: Sunyaev-Zel'dovich Effect},''
  \href{http://dx.doi.org/10.3847/1538-4357/aba694}{{\em Astrophys. J.}
  {\bfseries 900} no.~2, (2020) 110},
  \href{http://arxiv.org/abs/2003.06135}{{\ttfamily arXiv:2003.06135
  [astro-ph.CO]}}.

\bibitem{Ciuca:2017gca}
R.~Ciuca, O.~F. Hern\'andez, and M.~Wolman, ``{A Convolutional Neural Network
  For Cosmic String Detection in CMB Temperature Maps},''
  \href{http://dx.doi.org/10.1093/mnras/stz491}{{\em Mon. Not. Roy. Astron.
  Soc.} {\bfseries 485} (2019) 1377},
  \href{http://arxiv.org/abs/1708.08878}{{\ttfamily arXiv:1708.08878
  [astro-ph.CO]}}.

\bibitem{dumoulin2018guide}
V.~Dumoulin and F.~Visin, ``A guide to convolution arithmetic for deep
  learning,'' \href{http://arxiv.org/abs/1603.07285}{{\ttfamily
  arXiv:1603.07285 [stat.ML]}}.

\bibitem{ronneberger2015unet}
O.~Ronneberger, P.~Fischer, and T.~Brox, ``U-net: Convolutional networks for
  biomedical image segmentation,''
  \href{http://arxiv.org/abs/1505.04597}{{\ttfamily arXiv:1505.04597 [cs.CV]}}.

\bibitem{he2014convolutional}
K.~He and J.~Sun, ``Convolutional neural networks at constrained time cost,''
  \href{http://arxiv.org/abs/1412.1710}{{\ttfamily arXiv:1412.1710 [cs.CV]}}.

\bibitem{he2015deep}
K.~He, X.~Zhang, S.~Ren, and J.~Sun, ``Deep residual learning for image
  recognition,'' \href{http://arxiv.org/abs/1512.03385}{{\ttfamily
  arXiv:1512.03385 [cs.CV]}}.

\bibitem{srivastava2015highway}
R.~K. Srivastava, K.~Greff, and J.~Schmidhuber, ``Highway networks,''
  \href{http://arxiv.org/abs/1505.00387}{{\ttfamily arXiv:1505.00387 [cs.LG]}}.

\bibitem{he2016identity}
K.~He, X.~Zhang, S.~Ren, and J.~Sun, ``Identity mappings in deep residual
  networks,'' \href{http://arxiv.org/abs/1603.05027}{{\ttfamily
  arXiv:1603.05027 [cs.CV]}}.

\bibitem{balduzzi2018shattered}
D.~Balduzzi, M.~Frean, L.~Leary, J.~Lewis, K.~W.-D. Ma, and B.~McWilliams,
  ``The shattered gradients problem: If resnets are the answer, then what is
  the question?,'' \href{http://arxiv.org/abs/1702.08591}{{\ttfamily
  arXiv:1702.08591 [cs.NE]}}.

\bibitem{KayalibayJS17}
B.~Kayalibay, G.~Jensen, and P.~van~der Smagt, ``Cnn-based segmentation of
  medical imaging data,'' {\em CoRR} {\bfseries abs/1701.03056} (2017) ,
  \href{http://arxiv.org/abs/1701.03056}{{\ttfamily arXiv:1701.03056}}.
  \url{http://arxiv.org/abs/1701.03056}.

\bibitem{MilletariNA16}
F.~Milletari, N.~Navab, and S.~Ahmadi, ``V-net: Fully convolutional neural
  networks for volumetric medical image segmentation,'' {\em CoRR} {\bfseries
  abs/1606.04797} (2016) , \href{http://arxiv.org/abs/1606.04797}{{\ttfamily
  arXiv:1606.04797}}. \url{http://arxiv.org/abs/1606.04797}.

\bibitem{Zhang_2018}
Z.~Zhang, Q.~Liu, and Y.~Wang, ``Road extraction by deep residual u-net,''
  \href{http://dx.doi.org/10.1109/lgrs.2018.2802944}{{\em IEEE Geoscience and
  Remote Sensing Letters} {\bfseries 15} no.~5, (May, 2018) 749–753}.
  \url{http://dx.doi.org/10.1109/LGRS.2018.2802944}.

\bibitem{klambauer2017selfnormalizing}
G.~Klambauer, T.~Unterthiner, A.~Mayr, and S.~Hochreiter, ``Self-normalizing
  neural networks,'' \href{http://arxiv.org/abs/1706.02515}{{\ttfamily
  arXiv:1706.02515 [cs.LG]}}.

\bibitem{odena2016deconvolution}
A.~Odena, V.~Dumoulin, and C.~Olah, ``Deconvolution and checkerboard
  artifacts,'' \href{http://dx.doi.org/10.23915/distill.00003}{{\em Distill}
  (2016) }. \url{http://distill.pub/2016/deconv-checkerboard}.

\bibitem{Lewis:1999bs}
A.~Lewis, A.~Challinor, and A.~Lasenby, ``{Efficient computation of CMB
  anisotropies in closed FRW models},''
  \href{http://dx.doi.org/10.1086/309179}{{\em Astrophys. J.} {\bfseries 538}
  (2000) 473--476}, \href{http://arxiv.org/abs/astro-ph/9911177}{{\ttfamily
  arXiv:astro-ph/9911177}}.

\bibitem{Chardin:2019euc}
J.~Chardin, G.~Uhlrich, D.~Aubert, N.~Deparis, N.~Gillet, P.~Ocvirk, and
  J.~Lewis, ``{A deep learning model to emulate simulations of cosmic
  reionization},'' \href{http://dx.doi.org/10.1093/mnras/stz2605}{{\em Mon.
  Not. Roy. Astron. Soc.} {\bfseries 490} no.~1, (2019) 1055--1065},
  \href{http://arxiv.org/abs/1905.06958}{{\ttfamily arXiv:1905.06958
  [astro-ph.CO]}}.

\bibitem{kingma2014adam}
D.~P. Kingma and J.~Ba, ``Adam: A method for stochastic optimization,''
  \href{http://arxiv.org/abs/1412.6980}{{\ttfamily arXiv:1412.6980 [cs.LG]}}.

\bibitem{Meerburg:2017lfh}
P.~D. Meerburg, J.~Meyers, K.~M. Smith, and A.~van Engelen, ``{Reconstructing
  CMB fluctuations and the mean reionization optical depth},''
  \href{http://dx.doi.org/10.1103/PhysRevD.95.123538}{{\em Phys. Rev. D}
  {\bfseries 95} no.~12, (2017) 123538},
  \href{http://arxiv.org/abs/1701.06992}{{\ttfamily arXiv:1701.06992
  [astro-ph.CO]}}.

\bibitem{Ciuca:2019oka}
R.~Ciuca and O.~F. Hern\'andez, ``{Information Theoretic Bounds on Cosmic
  String Detection in CMB Maps with Noise},''
  \href{http://dx.doi.org/10.1093/mnras/stz3551}{{\em Mon. Not. Roy. Astron.
  Soc.} {\bfseries 492} no.~1, (2020) 1329--1334},
  \href{http://arxiv.org/abs/1911.06378}{{\ttfamily arXiv:1911.06378
  [astro-ph.CO]}}.

\bibitem{Feng:2018lwj}
C.~Feng and G.~Holder, ``{Detecting Electron Density Fluctuations from Cosmic
  Microwave Background Polarization using a Bispectrum Approach},''
  \href{http://dx.doi.org/10.1103/PhysRevD.97.123523}{{\em Phys. Rev. D}
  {\bfseries 97} no.~12, (2018) 123523},
  \href{http://arxiv.org/abs/1801.05396}{{\ttfamily arXiv:1801.05396
  [astro-ph.CO]}}.

\bibitem{Feng:2018eal}
C.~Feng and G.~Holder, ``{Searching for patchy reionization from cosmic
  microwave background with hybrid quadratic estimators},''
  \href{http://dx.doi.org/10.1103/PhysRevD.99.123502}{{\em Phys. Rev. D}
  {\bfseries 99} no.~12, (2019) 123502},
  \href{http://arxiv.org/abs/1808.01592}{{\ttfamily arXiv:1808.01592
  [astro-ph.CO]}}.

\bibitem{Hotinli:2020csk}
S.~C. Hotinli and M.~C. Johnson, ``{Reconstructing large scales at cosmic
  dawn},'' \href{http://arxiv.org/abs/2012.09851}{{\ttfamily arXiv:2012.09851
  [astro-ph.CO]}}.

\end{thebibliography}\endgroup

\end{document}